\documentstyle[11pt]{article}

\input amssym 
\catcode`\@=11\relax
\newwrite\@unused
\def\typeout#1{{\let\protect\string\immediate\write\@unused{#1}}}

\def\psglobal#1{
\immediate\special{ps: plotfile #1 }}
\def\psfiginit{\typeout{psfiginit}
\immediate\psglobal{figtex.pro}%
\special{ps:: /TeXMagnification {\the\mag} def}
}

\def\@nnil{\@nil}
\def\@empty{}
\def\@psdonoop#1\@@#2#3{}
\def\@psdo#1:=#2\do#3{\edef\@psdotmp{#2}\ifx\@psdotmp\@empty \else
    \expandafter\@psdoloop#2,\@nil,\@nil\@@#1{#3}\fi}
\def\@psdoloop#1,#2,#3\@@#4#5{\def#4{#1}\ifx #4\@nnil \else
       #5\def#4{#2}\ifx #4\@nnil \else#5\@ipsdoloop #3\@@#4{#5}\fi\fi}
\def\@ipsdoloop#1,#2\@@#3#4{\def#3{#1}\ifx #3\@nnil
       \let\@nextwhile=\@psdonoop \else
      #4\relax\let\@nextwhile=\@ipsdoloop\fi\@nextwhile#2\@@#3{#4}}
\def\@tpsdo#1:=#2\do#3{\xdef\@psdotmp{#2}\ifx\@psdotmp\@empty \else
    \@tpsdoloop#2\@nil\@nil\@@#1{#3}\fi}
\def\@tpsdoloop#1#2\@@#3#4{\def#3{#1}\ifx #3\@nnil
       \let\@nextwhile=\@psdonoop \else
      #4\relax\let\@nextwhile=\@tpsdoloop\fi\@nextwhile#2\@@#3{#4}}
\def\psdraft{
	\def\@psdraft{0}
	\def\@psdraftspecial{100}
}
\def\psdraftspecial{
	\def\@psdraft{0}
	\def\@psdraftspecial{0}
}
\def\psfull{
	\def\@psdraft{100}
}
\psfull

\newif\if@prologfile
\newif\if@postlogfile
\newif\if@bbllx
\newif\if@bblly
\newif\if@bburx
\newif\if@bbury
\newif\if@height
\newif\if@width
\newif\if@rheight
\newif\if@rwidth
\newif\if@clip
\newif\if@right
\newif\if@left
\newif\if@toplines
\newif\if@box
\newif\if@caption
\newif\if@surround
\newif\if@captionwidth
\newif\if@captionwrite
\newif\if@captionopen
\def\@p@@sclip#1{\@cliptrue}
\def\@p@@sfile#1{
		\def\@p@sfile{#1}
}
\def\@p@@sfigure#1{
		\def\@p@sfile{#1}
}
\def\@p@sfake{\hbox to 0pt{\hss Whatever\hss}}
\def\@p@@sbbllx#1{
		\@bbllxtrue
		\@d@mscratch=#1
		\edef\@p@sbbllx{\number\@d@mscratch}
}
\def\@p@@sbblly#1{
		\@bbllytrue
		\@d@mscratch=#1
		\edef\@p@sbblly{\number\@d@mscratch}
}
\def\@p@@sbburx#1{
		\@bburxtrue
		\@d@mscratch=#1
		\edef\@p@sbburx{\number\@d@mscratch}
}
\def\@p@@sbbury#1{
		\@bburytrue
		\@d@mscratch=#1
		\edef\@p@sbbury{\number\@d@mscratch}
}
\def\@p@@sheight#1{
		\@heighttrue
		\@d@mscratch=#1
   		\edef\@p@sheight{\number\@d@mscratch}
}
\def\@p@@swidth#1{
		\@widthtrue
		\@d@mscratch=#1
		\edef\@p@swidth{\number\@d@mscratch}
}
\def\@p@@srheight#1{
		\@rheighttrue
		\@d@mscratch=#1
		\edef\@p@srheight{\number\@d@mscratch}
}
\def\@p@@srwidth#1{
		\@rwidthtrue
		\@d@mscratch=#1
		\edef\@p@srwidth{\number\@d@mscratch}
}
\def\@p@@sright#1{\@righttrue \@surroundtrue}
\def\@p@@sleft#1{\@lefttrue \@surroundtrue}
\def\@p@@sextraheight#1{\@d@mextraheight=#1}
\def\@p@@sbox#1{\@boxtrue}
\def\@p@@scaption#1{\@captiontrue}
\def\@p@@stoplines#1{
		\@toplinestrue
		\@c@ttoplines=#1
}
\def\@p@@scaptionwidth#1{
		\@captionwidthtrue
	  	\@d@mcaptionwidth=#1
}
\def\@p@@scaptionwrite#1{
		\global\@captionwritetrue
		\global\@w@rname=\expandafter{\jobname_captions.tex}
		\typeout{Captions are written to \the\@w@rname}
}

\def\@p@@sprolog#1{\@prologfiletrue\def\@prologfileval{#1}}
\def\@p@@spostlog#1{\@postlogfiletrue\def\@postlogfileval{#1}}
\def\@cs@name#1{\csname #1\endcsname}
\def\@setparms#1=#2,{\@cs@name{@p@@s#1}{#2}}
\def\ps@init@parms{
		\@bbllxfalse \@bbllyfalse
		\@bburxfalse \@bburyfalse
		\@heightfalse \@widthfalse
		\@rheightfalse \@rwidthfalse
		\def\@p@sbbllx{}\def\@p@sbblly{}
		\def\@p@sbburx{}\def\@p@sbbury{}
		\def\@p@sheight{}\def\@p@swidth{}
		\def\@p@srheight{}\def\@p@srwidth{}
		\def\@p@sfile{}
		\def\@p@scost{10}
		\def\@sc{}
		\@prologfilefalse
		\@postlogfilefalse
		\@clipfalse
		\@rightfalse \@leftfalse
		\@boxfalse \@captionfalse
		\@toplinesfalse \@surroundfalse
		\@d@mextraheight=0pt
 		\@c@ttoplines=0
		\@pshape={} \def\@p@srheight@total{}
		\@captionwidthfalse \@d@mcaptionwidth=0pt
}
\def\parse@ps@parms#1{
	 	\@psdo\@psfiga:=#1\do
		   {\expandafter\@setparms\@psfiga,}}
\newif\ifno@bb
\newif\ifnot@eof
\newread\ps@stream
\newtoks\@linetok
\def\bb@missing{
	\typeout{psfig: searching \@p@sfile \space  for bounding box}
	\openin\ps@stream=\@p@sfile
	\no@bbtrue
	\not@eoftrue
	\catcode`\%=12
	\loop
		\read\ps@stream to \line@in
		\global\@linetok=\expandafter{\line@in}
		\ifeof\ps@stream \not@eoffalse \fi
		\@bbtest{\@linetok}
		\if@bbmatch\not@eoffalse\expandafter\bb@cull\the\@linetok\fi
	\ifnot@eof \repeat
	\catcode`\%=14
}	
\catcode`\%=12
\newif\if@bbmatch
\def\@bbtest#1{\expandafter\@a@\the#1
\long\def\@a@#1
     \ifx\@bbtest#2\@bbmatchfalse\else\@bbmatchtrue\fi}
\long\def\bb@cull#1 #2 #3 #4 #5 {
	\@d@mscratch=#2 bp\edef\@p@sbbllx{\number\@d@mscratch}
	\@d@mscratch=#3 bp\edef\@p@sbblly{\number\@d@mscratch}
	\@d@mscratch=#4 bp\edef\@p@sbburx{\number\@d@mscratch}
	\@d@mscratch=#5 bp\edef\@p@sbbury{\number\@d@mscratch}
	\no@bbfalse
}
\catcode`\%=14
\def\compute@bb{
		\no@bbfalse
		\if@bbllx \else \no@bbtrue \fi
		\if@bblly \else \no@bbtrue \fi
		\if@bburx \else \no@bbtrue \fi
		\if@bbury \else \no@bbtrue \fi
		\ifno@bb \bb@missing \fi
		\ifno@bb \typeout{FATAL ERROR: no bb supplied or found}
			\no-bb-error
		\fi
		\count203=\@p@sbburx
		\count204=\@p@sbbury
		\advance\count203 by -\@p@sbbllx
		\advance\count204 by -\@p@sbblly
		\edef\@bbw{\number\count203}
		\edef\@bbh{\number\count204}
}
\def\in@hundreds#1#2#3{\count240=#2 \count241=#3
		     \count100=\count240	
		     \divide\count100 by \count241
		     \count101=\count100
		     \multiply\count101 by \count241
		     \advance\count240 by -\count101
		     \multiply\count240 by 10
		     \count101=\count240	
		     \divide\count101 by \count241
		     \count102=\count101
		     \multiply\count102 by \count241
		     \advance\count240 by -\count102
		     \multiply\count240 by 10
		     \count102=\count240	
		     \divide\count102 by \count241
		     \count200=#1\count205=0
		     \count201=\count200
			\multiply\count201 by \count100
		     	\advance\count205 by \count201
		     \count201=\count200
			\divide\count201 by 10
		     	\multiply\count201 by \count101
			\advance\count205 by \count201
		     \count201=\count200
			\divide\count201 by 100
			\multiply\count201 by \count102
			\advance\count205 by \count201
		     \edef\@result{\number\count205}
}
\def\compute@wfromh{
		\in@hundreds{\@p@sheight}{\@bbw}{\@bbh}
		\edef\@p@swidth{\@result}
}
\def\compute@hfromw{
		\in@hundreds{\@p@swidth}{\@bbh}{\@bbw}
		\edef\@p@sheight{\@result}
}
\def\compute@handw{
		\if@height
			\if@width
			\else
				\compute@wfromh
			\fi
		\else
			\if@width
				\compute@hfromw
			\else
				\edef\@p@sheight{\@bbh}
				\edef\@p@swidth{\@bbw}
			\fi
		\fi
}
\def\compute@resv{
		\if@rheight \else \edef\@p@srheight{\@p@sheight} \fi
		\if@rwidth \else \edef\@p@srwidth{\@p@swidth} \fi
		\edef\@p@srheight@total{\@p@srheight}
}
\newtoks\@pshape
\def\@c@ttoplines{\count120}
\def\@c@theightcount{\count121}
\def\@c@tshapecount{\count122}
\newdimen\@d@mwidthshape
\newdimen\@d@mextraheight
\newdimen\@d@mscratch
\def\compute@parshape{
	\if@right
		\if@left
	   		\typeout{error: Can't have both left and right set}
			\@leftfalse
		\fi
	\fi
	\@d@mscratch=\@p@swidth truesp
	\divide \@d@mscratch by 19
	\multiply \@d@mscratch by 20
	\edef\@p@swidthdimen{\the\@d@mscratch}
	\@c@tshapecount=\@c@ttoplines
 	\@d@mscratch=\baselineskip
	\multiply \@d@mscratch by \@c@ttoplines
	\advance \@d@mscratch by .4\baselineskip
    	\edef\@p@stopdistance{\the\@d@mscratch }
	\@d@mscratch=\@p@sheight truesp
	\divide \@d@mscratch by 2
	\edef\@p@shalfboxheight{\the\@d@mscratch}
	\if@toplines
		\loop \@pshape=\expandafter{\the\@pshape 0pt \hsize}
		\advance\@c@ttoplines by -1
		\ifnum\@c@ttoplines>0 \repeat
	\fi
   	\@c@theightcount=\@p@srheight@total
	\advance \@c@theightcount by \@d@mextraheight
	\divide  \@c@theightcount by \baselineskip
	\advance \@c@theightcount by 1
    	\advance \@c@tshapecount by \@c@theightcount
	\advance \@c@theightcount by -1
	\@d@mwidthshape=\hsize
     	\advance \@d@mwidthshape by -\@p@swidthdimen
	\if@left
		\def\@moveshape{0pt}
		\ifnum\@c@theightcount>0
		      	\loop
			\@pshape=%
			\expandafter{\the\@pshape %
					\@p@swidthdimen \@d@mwidthshape}
			\advance \@c@theightcount by -1
			\ifnum\@c@theightcount>0 \repeat
		\else
			\advance \@c@tshapecount by 1
		\fi
	\fi
	\if@right
		\@d@mscratch=\hsize
		\advance \@d@mscratch by -\@p@swidth truesp
		\edef\@moveshape{\@d@mscratch}
		\ifnum\@c@theightcount>0
			\loop
			\@pshape=\expandafter{\the\@pshape 0pt \@d@mwidthshape}
			\advance \@c@theightcount by -1
			\ifnum\@c@theightcount>0 \repeat
		\else
			\advance \@c@tshapecount by 1
		\fi
	\fi
	\ifnum \@p@srheight=0
		\@pshape={}
		\@c@tshapecount = 0
	\else
	 	\@pshape=\expandafter{\the\@pshape 0pt \hsize}
	\fi
}
\def\@p@ssetsurroundboxes{
	\global\parshape=\count122 \the\@pshape		
 	\moveright\@moveshape
	\vbox to 0pt\bgroup\hskip0pt\vskip\@p@stopdistance
}
\newtoks\@captiontok
\newbox\@b@xcaption
\newdimen\@d@mcaptionwidth
\newdimen\@d@mcaptionheight
\newwrite\@w@rcaption
\newtoks\@w@rname
\def\setcaption#1{\@captiontok={#1}}
\def\@set@caption{
	\setbox\@b@xcaption\vbox{\hsize\@d@mcaptionwidth
	\tolerance=9000 \baselineskip=11.4pt
	\noindent\relax\the\@captiontok}
	\if@captionwrite
		\if@captionopen
		\else
			\global\@captionopentrue
			\immediate\openout\@w@rcaption=\the\@w@rname
		\fi
		\immediate\write\@w@rcaption{\the\@captiontok}
		\immediate\write\@w@rcaption{}
	\fi
}
\def\compute@caption{
	\if@captionwidth
	\else
		\@d@mcaptionwidth = \@p@swidth truesp
		\divide \@d@mcaptionwidth by 20
		\multiply \@d@mcaptionwidth by 17
	\fi
	\@set@caption
	\@d@mcaptionheight=\ht\@b@xcaption
	\if@rheight
	\else
		\count100=\@p@srheight
	   	\advance \count100 by \@d@mcaptionheight
	   	\advance \count100 by \bigskipamount
	   	\advance \count100 by \medskipamount
	   	\edef\@p@srheight@total{\number\count100}
	\fi
}
\newif\if@alreadyjtem \@alreadyjtemfalse
\def\newpar{\hfil\vadjust{\vskip\parskip}%
	{\count100=\parskip
	\count101=\baselineskip
	\divide\count101 by 10  \multiply\count101 by 3
	\advance \count100 by \count101
	\divide\count100 by \baselineskip
	\advance\count100 by \prevgraf
	\global\prevgraf=\count100}%
	\break\if@alreadyjtem\else\indent\fi%
}
\let\sav@par=\par
\def\jtem#1{%
    	\if@alreadyjtem\else\bgroup\fi
	\def\par{\sav@par\egroup\sav@par}
	\if@alreadyjtem\else\leftskip\parindent\fi
	\@alreadyjtemtrue
	\noindent\hskip0pt
	\llap{#1\ }\ignorespaces
}
\def\compute@sizes{%
	\compute@bb
	\compute@handw
  	\compute@resv
	\if@caption
		\compute@caption
	\fi
	\if@surround
		\compute@parshape
	\fi
}
\def\@p@sdospecials{
	\ifnum\@p@scost<\@psdraft
	       	\typeout{psfig: including \@p@sfile \space }
	\fi
	\special{ps::[begin] 	\@p@swidth \space \@p@sheight \space
			\@p@sbbllx \space \@p@sbblly \space
			\@p@sbburx \space \@p@sbbury \space
			startTexFig \space }
	\ifnum\@p@scost<\@psdraft
		\if@clip
			\typeout{(clip)}
			\special{ps:: \@p@sbbllx \space \@p@sbblly \space
				\@p@sbburx \space \@p@sbbury \space
			    	doclip \space }
		\fi
	\fi
	\if@box
		\typeout{(box)}
  		\special{ps:: \@p@sbbllx \space \@p@sbblly \space
			\@p@sbburx \space \@p@sbbury \space
		    	dobox \space }
	\fi
	\ifnum\@p@scost<\@psdraft
		\if@prologfile
	    		\special{ps: plotfile \@prologfileval \space }
		\fi
		\special{ps: plotfile \@p@sfile \space }
    		\if@postlogfile
			\special{ps: plotfile \@postlogfileval \space }
		\fi
	\fi
	\special{ps::[end] endTexFig \space }
}
\newif\if@putinvbox

\def\psfig#1{{%
	\ifhmode%
		\vbox\bgroup
		\@putinvboxtrue
	\else
		\@putinvboxfalse
	\fi
       	\ps@init@parms
	\parse@ps@parms{#1}
       	\compute@sizes
	\if@surround
		\psfig@for@surround
	\else
		\psfig@for@regular
	\fi
	\if@putinvbox
       		\egroup
	\fi
}}
\def\psfig@for@surround{%
	\@p@ssetsurroundboxes
	\ifnum\@p@scost<\@psdraft
		\@p@sdospecials
		\vbox to \@p@srheight true sp{\vss}
       	\else
		\if@box
			\@p@sdospecials
		\fi
		\vbox to \@p@srheight true sp{
			\vskip\@p@shalfboxheight
			\hbox to \@p@srwidth true sp{
				\hss
				\ifnum\@p@scost<\@psdraftspecial
					\@p@sfile
				\else
					\@p@sfake
				\fi
      				\hss
			}
		\vss
		}
	\fi
	\if@caption
		\medskip
		\hbox to \@p@srwidth true sp{
			\hss
			\box\@b@xcaption
			\hss
		}
 		\medskip
	\fi
	\vss\egroup
	\vskip-\parskip
}

\def\psfig@for@regular{%
	\if@putinvbox
	\else
		\vskip\parskip
	\fi
	\ifnum\@p@scost<\@psdraft
		\@p@sdospecials
		\vbox to \@p@srheight true sp{%
			\hbox to \@p@srwidth true sp{
			\hfil
			}
		\vfil
		}
       	\else
		\if@box
			\@p@sdospecials
		\fi
	    	\vbox to \@p@srheight true sp{
			\vss
			\hbox to \@p@srwidth true sp{
				\hss
				\ifnum\@p@scost<\@psdraftspecial
					\@p@sfile
				\else
					\@p@sfake
				\fi
				\hss
			}
		    	\vss
		}
	\fi
	\if@caption
		\medskip
		\hbox to \@p@srwidth true sp{
			\hss
			\box\@b@xcaption
			\hss
		}
		\bigskip
	\fi
	\if@putinvbox
	\else
		\vskip-\parskip
	\fi
}
\catcode`\@=12\relax
\psfiginit

 1
\font\subscriptsizebbfont=msbm7 scaled \magstep 1
 1
 3
\font\bbfont=msbm10 scaled \magstep1  

\def\subscriptsizeBbb#1{\hbox{\subscriptsizebbfont #1}}

\def\Bbb#1{\hbox{\bbfont #1}}

\newcommand{\pr}{\mbox{\rm pr}}

\begin{document}

\begin{titlepage}

$ $

\vspace{-1.5cm} 

\noindent\hspace{-1cm}
\parbox{6cm}{\footnotesize VAX/VMS: March 1997 \newline
 Micron MPC P166: July 1997 }\
  \hspace{7.5cm}\
  \parbox{5cm}{{\sc umtg} -- {\small 192}   \newline
               {\tt hep-th/9707196}  }

\vspace{1.8cm}

\centerline{\large\bf Remarks on the Geometry of Wick Rotation in QFT}
\vspace{1ex}
\centerline{\large\bf and its Localization on Manifolds}

\vspace{1.5cm}

\centerline{\large Chien-Hao Liu\footnote{
 Address after August 1, 1997:
   Department of Mathematics, University of Texas at Austin,
   Austin, Texas 78712.
 E-mail: chienliu@phyvax.ir.miami.edu
                  /(after August 20, 1997) chienliu@math.utexas.edu}}

\vspace{1.1em} 

\centerline{\it Department of Physics, University of Miami}
\centerline{\it P.O.\ Box 248046, Coral Gables, FL.\ 33124}

\vspace{2em}

\begin{quotation}
\centerline{\bf Abstract}
\vspace{0.3cm}

\baselineskip 12 pt  

{\small
The geometric aspect of Wick rotation in quantum field theory and its
localization on manifolds are explored. After the explanation of the
notion and its related geometric objects, we study the topology of
the set of landing $W$ for Wick rotations and its natural stratification.
These structures in two, three, and four dimensions are computed
explicitly. We then focus on more details in two dimensions.
In particular, we study the embedding of $W$ in the ambient space of
Wick rotations, the resolution of the generic metric singularities of a
Lorentzian surface $\Sigma$ by local Wick rotations, and some related
$S^1$-bundles over $\Sigma$.
} 
\end{quotation}

\bigskip

\baselineskip 12pt

{\footnotesize
\noindent
{\bf Key words:} \parbox[t]{12cm}{
Quantum field theory, Wick rotation, the set of landing,
 Milnor fibration, \newline
local Wick rotation, stratification, $A_1$-singularity. 
}} 

\bigskip

\noindent
MSC number 1991: 15A63, 22E70, 57R22, 81T99; 51P05, 58A35.

\bigskip

\baselineskip 11pt  

{\footnotesize
\noindent{\bf Acknowledgements.}
This is my last work at U.M., completed just before a new starting point
in my life.
It is beyond words for my gratitude to my physics advisor
 Orlando Alvarez for his crucial teachings and guidances in my study,
 his patience in listening to me, and his encouragements and friendship
 both in work and life since the Berkeley period.
I would like to thank also my two other advisors: Hai-Chau Chang,
 William Thurston, and friends: Hung-Wen Chang, Ying-Ing Lee,
 who have influenced me in many aspects along the years.
Many thanks are to
 O.A.\ Ching-Li Chai, H.-W.C.\ for discussions and techenical helps on
 this work;
 (time-ordered) David Gross, Korkut Bardak\c{c}i, O.A.,
  Thomas Curtright, Rafael Nepomechie for courses on QFT that
  motivate it; and
 H.-W.C., Chong-Sun Chu, Pei-Ming Ho for many discussions and helps in
  learning QFT.
Tremendous gratitude is also to
  T.C., R.N.\ for their generosity of allowing me to knock at their
 doors whenever in need; the theory group:
  O.A., T.C., Luca Mezincescu, James Nearing, R.N., Lev Rozansky,
  Radu Tartar, Paul Watts for enrichment of knowledge;
 Marco Monti for assistance in VAX/VMS, Xfig, and setting up a
  self-contained PC for the work;
 the chair George Alexandrakis for his kindness, generosity, and
  great understanding that have eased my life
 and the staff: Lourdes Castro, Judy Mallery, Melody Pastore
  for many conveniences provided during my stay at U.M., without which
  no works would be possible;
and O.A., T.C., Martin Halpern, R.N., Lorenzo Sadun, W.T.,
 Neil Turok for creating a new future for me.
Behind all these I thank, again beyond any words,
 my parents and Ling-Miao for their moral supports, patience, and
 understandings that I can never reward them.
} 

\end{titlepage}


$ $

\vspace{-4em}
\centerline{\sc Geometry and Localization of Wick Rotations in QFT}
\vspace{4em}

{\small
\baselineskip 11pt  
\noindent
{\bf Contents.}
\begin{quote}
  \hspace{1em} Introduction.

  1. Wick rotation and its localization.
  \vspace{-1ex}
  \begin{quote}
   \raisebox{.2ex}{\scriptsize $\bullet$}
    Defintion and basic geometric objects involved.

   \raisebox{.2ex}{\scriptsize $\bullet$}
   Generalization to manifolds.
  \end{quote}

  2. The set of landing $W$ for Wick rotations.
  \vspace{-1ex}
  \begin{quote}
   \raisebox{.2ex}{\scriptsize $\bullet$}
   The topology of $W$.

   \raisebox{.2ex}{\scriptsize $\bullet$}
   A natural stratification of $\overline{W}$.
  \end{quote}

  3. $\overline{W}$ in two, three, and four dimensions.

  4. Wick rotations in two dimensions.
  \vspace{-1ex}
  \begin{quote}
   \raisebox{.2ex}{\scriptsize $\bullet$}
   How $W(2)$ embeds in $M(2,{\Bbb C})$.

   \raisebox{.2ex}{\scriptsize $\bullet$}
   Local Wick rotations of a surface.
  \end{quote}
\end{quote}
} 

\bigskip

\baselineskip 14pt 

\begin{flushleft}
{\Large\bf Introduction.}
\end{flushleft}
Quantum field theory (QFT) mingles together bold ideas, theoretical
concepts, and tricks for practical computations. {\it Wick rotation}
is one such example. In the computation of Feynman diagrams, one
constantly faces integrals of meromorphic functions over the
Minkowskian space-time or momentum $4$-space, which are literally
divergent. However by pushing the real time axis into the complex time
plane and rotate it to the purely imaginary time axis
({\sc Figure I-1}), the metric becomes Euclidean and the integrals
become ones over Euclidean $4$-space. One can then introduce either
cutoffs or dimensional regularization to make these integrals finite
and next tries to understand the divergent behavior when letting the
cutoffs go to infinity, the dimensions of space-time approach $4$, or the
imaginary time axis rotated back to the real one. Indeed in Wilson's
renormalization theory, the attitude of taking the original theory on
the Minkowskian space-time as the limit of analytic continuations (via
inverse Wick rotation) from the theory on the Euclidean $4$-space
becomes more profound since the notion of scales and cutoffs is
fundamental in his theory and only in Euclidean space do they have
better control of (components of) momenta. Due to this origin, the
analytic aspect of Wick rotation has been discussed in the literature
and the notion has become a standard one in QFT
(e.g.\ [G-J], [I-Z], [P-S]). In contrast, the geometric aspect of it
seems dimmed, due to lack of necessity.

\begin{figure}[htbp]
\setcaption{{\sc Figure I-1.}
\baselineskip 14 pt  
    Wick rotation in QFT is a rotation of
    the real time axis to the imaginary time axis in the complex-time
    plane.}
\centerline{
 \psfig{figure=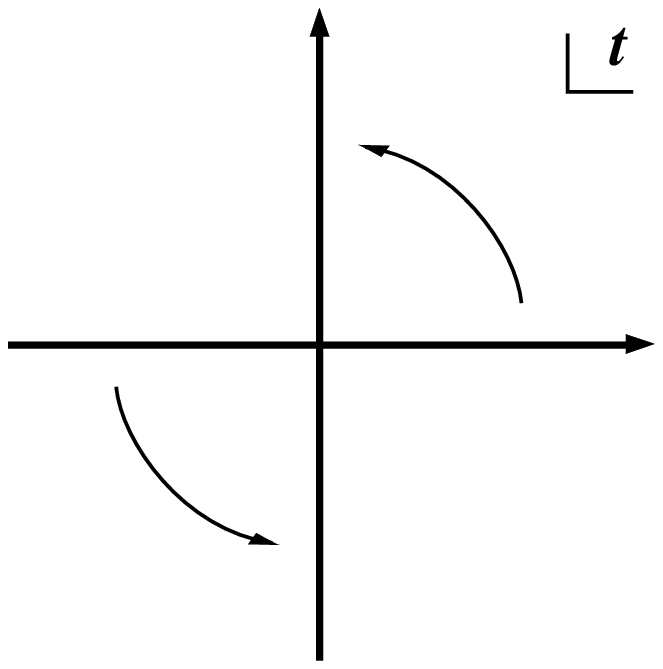,width=13cm,caption=}}
\end{figure}

However, the study of gravity and string theory forces one to consider
manifolds and press on, asking:
{\it How indeed are Lorentzian manifolds Wick-rotated to Riemannian
manifolds?} and {\it How such rotations relate the Minkowskian and
the Euclidean theory?} It is for such harder questions that the
geometric aspect of Wick rotation should demonstrate its roles.

In this paper, we shall focus on Wick rotation as a geometric procedure
that relates metrics of different signatures and study the geometry
behind this in its own right. After giving the general definition
of Wick rotation for a vector space motivated from above and its local
generalization to manifolds in Sec.\ 1, we study the structure of the
set of landing $W$ for Wick rotations in Sec.\ 2. This set consists of
the terminal vector spaces of Wick rotations, on which the metric is
required to resume real-valued so that they become the usual real inner
product spaces again. The different signatures of the inner product on
the terminal spaces that Wick-rotations can lead to induce a
stratification of $W$. After the general study of these in Sec.\ 2, we
compute them explicitly for the cases of two, three, and four demensions
in Sec.\ 3. We then turn our focus to more details of Wick rotations in
two dimensions in Sec.\ 4. We study the embedding of $W$ in the ambient
space of Wick rotations, the resolution of the generic metric
singularities of a Lorentzian surface $\Sigma$ by local Wick rotations,
and some related $S^1$-bundles over $\Sigma$.

\bigskip

\section{Wick rotation and its localization.}

In this section we shall formulate a defintion of Wick rotations
directly from QFT and discuss its generalizations to manifolds. Readers
are referred to [Mi] and [Di2] for some terminology and background from
the theory of singularities.

\bigskip

\begin{flushleft}
{\bf Definition and basic geometric objects involved.}
\end{flushleft}
The reflection on Wick rotation in QFT motivates us the following
general definition of Wick rotations.

\bigskip

\noindent{\bf Definition 1.1 [Wick rotation].}\hspace{.6ex}
 Let $V$ be a real vector space with a non-degenerate inner product
 $\rho$ and $V_{\subscriptsizeBbb C}=V\otimes {\Bbb C}$ be its
 complexification with the natural ${\Bbb C}$-linear inner product
 $\rho_{\subscriptsizeBbb C}=\rho\otimes{\Bbb C}$.
 Let $\iota:V\hookrightarrow V_{\subscriptsizeBbb C}$ be the natural
 inclusion. Then a {\it Wick rotation} of $V$ is a smooth family of
 injective ${\Bbb R}$-linear homomorphisms
 $$
  f_t\;:\; V\; \longrightarrow \; V_{\subscriptsizeBbb C}
  \hspace{1em}\mbox{with $t\in [0,1]$}\,,
 $$
 such that $f_0=\iota$ and that $\rho_t$ defined by $f_t^{\ast}\rho$ is
 non-degenerate on $V$ for all $t$ with $\rho_1$ real-valued.

\bigskip

Let $\mbox{\it Hom}\,_{\subscriptsizeBbb R}(V,V_{\subscriptsizeBbb C})$
be the space of all ${\Bbb R}$-linear homomorphisms from $V$ to
$V_{\subscriptsizeBbb C}$. There are two subsets in
$\mbox{\it Hom}\,_{\subscriptsizeBbb R}(V,V_{\subscriptsizeBbb C})$
that are particularly important in the geometric picture of Wick
rotations:
\begin{quote}
 \hspace{-2em}
 (1) the {\it set $W$ of landing} that contains all $f$ in
     $\mbox{\it Hom}\,_{\subscriptsizeBbb R}(V,V_{\subscriptsizeBbb C})$
     with $f^{\ast}\rho$ real-valued, and

 \hspace{-2em}
 (2) the {\it set $\Xi$ of degeneracy} that contains all $f$ in
     $\mbox{\it Hom}\,_{\subscriptsizeBbb R}(V,V_{\subscriptsizeBbb C})$
     with $f^{\ast}\rho$ degenerate.
\end{quote}
Both of them are cones in
$\mbox{\it Hom}\,_{\subscriptsizeBbb R}(V,V_{\subscriptsizeBbb C})$
with apex the zero map $O$. In terms of them, a Wick rotation is simply
a path in the complement
$\mbox{\it Hom}\,_{\subscriptsizeBbb R}(V,V_{\subscriptsizeBbb C})-\Xi$
that begins with $\iota$ and lands on $W$.

Notice that 
$\mbox{\it Hom}\,_{\subscriptsizeBbb R}(V,V_{\subscriptsizeBbb C})$ can
be identified with the space $M(n,{\Bbb C})$ of $n\times n$ matrices with
complex entries (or of $n$-tuples of vectors in ${\Bbb C}^n$) once a basis
of $V$ is chosen. Also observe that, given $n$ vectors
$\xi_1,\ldots,\xi_n$ in $V_{\subscriptsizeBbb C}$ with each vector
regarded as a column vector, one has the identity
$$
 \det\left[\rule{0ex}{2ex}
   \rho_{\subscriptsizeBbb C}(\xi_i,\xi_j)\right]_{ij} \;
 =\;
\left(\det\left[\rule{0ex}{2ex}\xi_1,\cdots,\xi_n\right]\right)^2\,,
$$
which follows from the fact that any $n\times n$ matrix is conjugate by
an elememt in $\mbox{\it SL}\,(n,{\Bbb C})$ to the Jordan form, for which
the above identity holds by induction on the rank of elementary Jordan
matrices. This implies that the set $\Xi$ is simply the variety in
${\Bbb C}^{n^2}$ described by
$\det\left[\rule{0ex}{2ex}\xi_1,\cdots,\xi_n\right]=0$. Thus it is known
([Di1-2]) that $\Xi$ can be naturally stratified by the subset
$\Xi_{(r)}$ that consists of
$\left[\rule{0ex}{2ex}\xi_1,\cdots,\xi_n\right]$ of rank $r$. Each
$\Xi_{(r)}$ is a smooth connected manifold of dimension $n^2-(n-r)^2$.
The singular set of $\Xi$ is the union $\cup_{r=0}^{n-2}\,\Xi_{(r)}$,
which has codimension $3$ in $\Xi$. The complement of $\Xi$ in
$M(n,{\Bbb C})$ can now be identified with the general linear group
$\mbox{\it GL}\,(n,{\Bbb C})$. Since $\mbox{\it GL}\,(n,{\Bbb C})$ is
path-connected, no matter how two inner products are dressed to $V$,
they can always be connected to each other by a Wick rotation.

Let $S^{2n^2-1}$ be any $(2n^2-1)$-sphere in $M(n,{\Bbb C})$ that bounds the
zero matrix $O_n$ and is transverse to all the rays from $O_n$. Since $\Xi$
and $W$ are cones at $O_n$, to understand them, one may as well study their
intersection $K_{\Xi}$ and $K_W$ with $S^{2n^2-1}$. Since
$\mbox{\it GL}\,(n,{\Bbb C})
  =\mbox{\it SL}\,(n,{\Bbb C})\times{\Bbb C}^{\ast}$, where
${\Bbb C}^{\ast}={\Bbb C}-\{0\}$, and $\det$ is homogeneous, the map
$$
 \varphi\: :\: S^{2n^2-1} - K_{\Xi}\: \longrightarrow\: S^1 
$$
defined by $\varphi=\det/\mbox{\raisebox{-.3ex}{$|\det|$}}$ gives a
trivial (Milnor) fibration with (Milnor) fiber
$\mbox{\it SL}\,(n,{\Bbb C})$. And $K_W-K_{\Xi}$ lies in
$\varphi^{-1}(\{1,i,-1,-i\})$, a union of four fibers.
(Cf.\ {\sc Figure} 4-4(b).) Consequently, for a physical quantity
${\cal F}$ that depends analytically on the inner product, its extension
along different Wick rotations from a given inner product to another may
assume different values. On the other hand, since
$$
 \pi_1(\mbox{\it SL}\,(n,{\Bbb C}))\;
 =\; \pi_2(\mbox{\it SL}\,(n,{\Bbb C}))\; =\;0\,,
$$
a fact that follows either from the property of the Milnor fiber of a
normal hypersurface singularity like $\Xi$ ([Di2]) or from some direct
argument, besides the set of poles of ${\cal F}$ in
$\mbox{\it GL}\,(n,{\Bbb C})$, the factor $S^1$, which may be regarded
as the generator of $\pi_1(\mbox{\it GL}\,(n,{\Bbb C}))$ is the only
complication to the analytic extension of ${\cal F}$ from the global
topology of $\mbox{\it GL}\,(n,{\Bbb C})$.

\bigskip

\noindent
{\it Remark 1.2.} In retrospect, one may define a Wick rotation
 more restrictively as a path that lies in a one-parameter subgroup
 of $\mbox{\it GL}\,(n,{\Bbb C})$ with one end on the identity and
 the other end on $W$. This is indeed what happens in QFT.

\bigskip

\begin{flushleft}
{\bf Generalization to manifolds.}
\end{flushleft}
When moving on to a manifold $M=(M,\rho)$, where $\rho$ is a
non-degenerate metric on $M$, the notion of Wick rotations of a vector
space indeed can be generalized in at least two ways:

\bigskip

\noindent
{\bf Definition 1.3 [global and local Wick rotation].}\hspace{.6ex}
(1) {\it Global} : Let $M_{\subscriptsizeBbb C}$ be a complex manifold
 that admits an embedding
 $\iota: M\hookrightarrow M_{\subscriptsizeBbb C}$ as a totally real
 submanifold and a non-degenerate ${\Bbb C}$-bilinear metric tensor
 $\rho_{\subscriptsizeBbb C}$ such that, when restricted to
 $T|_M M_{\subscriptsizeBbb C}$, $\rho_{\subscriptsizeBbb C}$ is simply
 the complexification $\rho\otimes{\Bbb C}$.
 A {\it global Wick rotation} is then a smooth family of embeddings
 (or immersions)
 $f_t: M \rightarrow M_{\subscriptsizeBbb C}$, $t\in[0,1]$, such that
 $f_0=\iota$ and that $f_t^{\ast}\rho_{\subscriptsizeBbb C}$ is a
 non-degenerate (complex-valued) inner product tensor on $M$ for all
 $t$ with $f_1^{\ast}\rho_{\subscriptsizeBbb C}$ real-valued.

\medskip

\noindent
(2) {\it Local} : QFT teaches us that whenever there is a notion defined
 on a vector space, one can consider its generalization to a vector bundle
 by {\it localizing} the notion. Thus one considers
 $T_{\subscriptsizeBbb C}M$, the complexification
 $T_{\ast}M\otimes{\Bbb C}$ of the tangent bundle, with the
 inner product tensor $\rho_{\subscriptsizeBbb C}=\rho\otimes{\Bbb C}$.
 Let $\iota: T_{\ast}M\rightarrow T_{\subscriptsizeBbb C}M$ be the
 natural bundle inclusion. A {\it local} (or, perhaps better,
 {\it fiberwise}) {\it Wick rotation} is then a smooth family of
 injective bundle homomorphisms
 $f_t: T_{\ast}M \rightarrow T_{\subscriptsizeBbb C}M$, $t\in [0,1]$,
 whose induced maps on $M$ are the identity map, such that
 $f_0=\iota$ and $f_t^{\ast}\rho_{\subscriptsizeBbb C}$ is
 non-degenerate along each fiber of $T_{\subscriptsizeBbb C}M$ for all
 $t$ with $f_1^{\ast}\rho_{\subscriptsizeBbb C}$ real-valued.
 ({\sc Figure 1-1.})

\begin{figure}[htbp]
\setcaption{{\sc Figure 1-1.}
\baselineskip 14pt 
  Wick rotation can be localized on a
  manifold $M$. They are fiberwise ``rotations" of $T_{\ast}M$ in its
  own complexification $T_{\subscriptsizeBbb C}M$, subject to some
  non-degeneracy requirement. In this picture, only one fiber of
  $T_{\subscriptsizeBbb C}M$ is shown.   }
\centerline{\psfig{figure=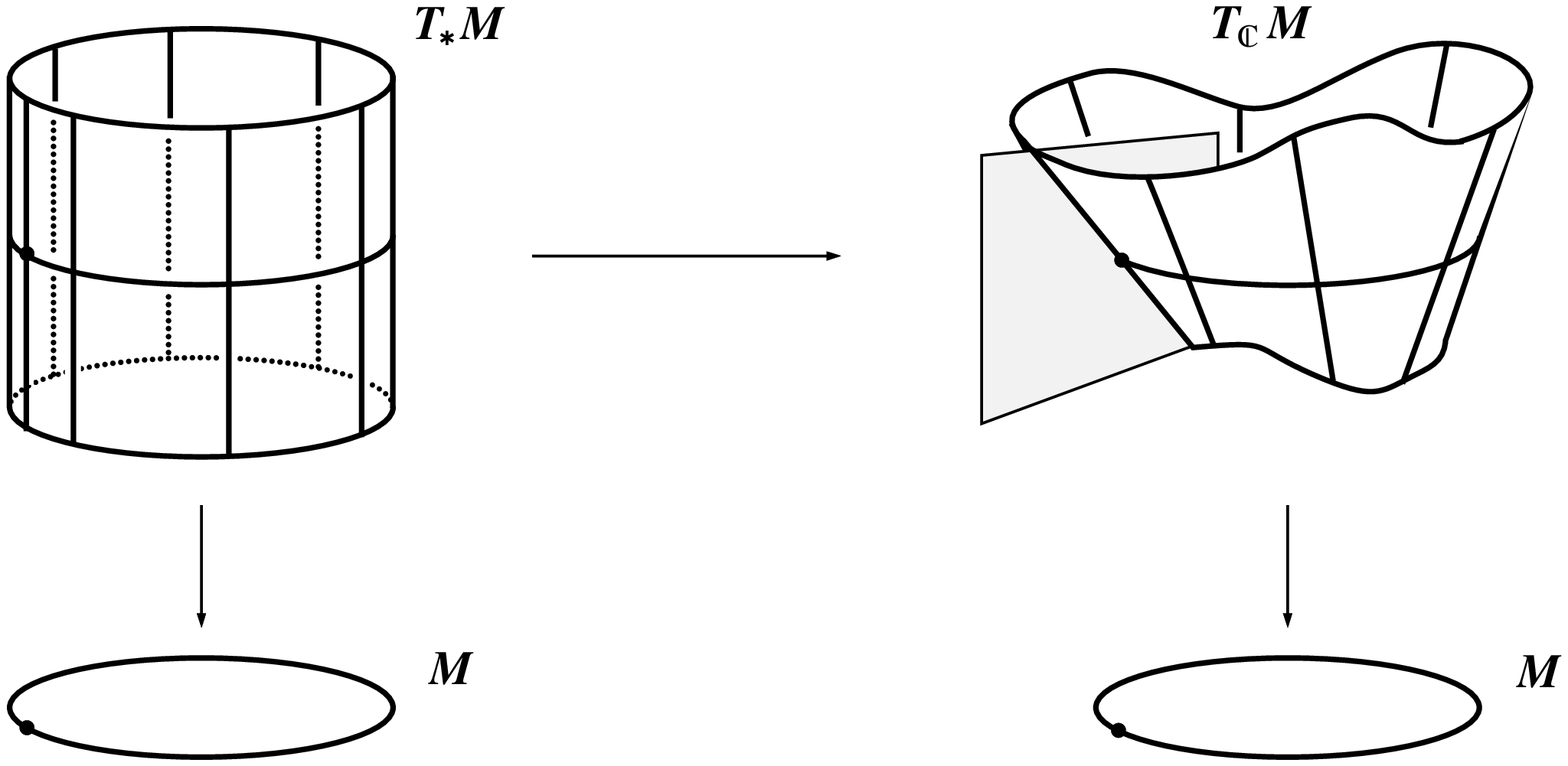,width=13cm,caption=}}
\end{figure}

\bigskip

\noindent
{\it Remark 1.4.} There is a subtlety in the above definition. If $M$
 admits a non-degenerate metric, then the type $(p,q)$ of the metric that
 encodes the dimension of a maximal positive-definite and a maximal
 negative-definite subspace in each tangent space to $M$ is constant. This
 can happen if and only if $M$ admits a nowhere degenerate $p$-plane field
 and, hence, puts a constraint on the topology of $M$. In general, one may
 need to relax the requirement of non-degeneracy of
 $f_1^{\ast}\rho_{\subscriptsizeBbb C}$ everythere and allow generic metric
 singularities to occur, e.g.\ a discrete set for
 $f_1^{\ast}\rho_{\subscriptsizeBbb C}$ Lorentzian. Under this relaxation,
 $f_1^{\ast}\rho_{\subscriptsizeBbb C}$ may indeed be a
 {\it heterotic metric} on $M$. The set of metric degeneracy could
 decompose $M$ into a collection of domains, with the metric restricted
 to each domain having a different type.

\bigskip

In the following sections, we shall try to understand $W$ better and
shall discuss also local Wick rotations for surfaces. Readers note that
the identifications
$\mbox{\it Hom}\,_{\subscriptsizeBbb R}(V,V_{\subscriptsizeBbb C})
  = M(n,{\Bbb C})$,
$V_{\subscriptsizeBbb C}={\Bbb C}^n={\Bbb R}^{2n}$, and vectors as
column vectors, when a basis is chosen, are implicit in many places;
$A^t$ means the transpose of a matrix $A$. An {\it $n$-frame} in this
paper will mean an {\it ${\Bbb R}$-linearly independent} $n$-tuple of
vectors (or vector fields).

\bigskip

\section{The set $W$ of landing for Wick rotations.}

Since Wick rotations happpn in the complement of $\Xi$ in
$M(n,{\Bbb C})$, it is the intersection
$W\cap \mbox{\it GL}\,(n,{\Bbb C})$ that comes into play. However, it
turns out more convenient to study the larger open dense subset in $W$
that consists of all $n$-frames. For simplicity of notation, we shall
denote the latter still by $W$.

\bigskip

\begin{flushleft}
{\bf The topology of $W$.}
\end{flushleft}
Let $G_n({\Bbb R}^{2n})$ be the Grassmann manifold of $n$-dimensional
subspaces in ${\Bbb R}^{2n}$. Recall that there is a tautological
$\mbox{\it GL}\,(n,{\Bbb R})$-bundle over $G_n({\Bbb R}^{2n})$. In
terms of this, $W$ can be realized as the restriction of this bundle
to the subset $\overline{W}$ in $G_n({\Bbb R}^{2n})$ that consists of
all the real $n$-dimensional subspaces $E$ in $V_{\subscriptsizeBbb C}$
with $\rho_{\subscriptsizeBbb C}|_E$ real. Our task now is to understand
this base space $\overline{W}$.

Let ${\Bbb C}^{\ast}={\Bbb C}-\{0\}$. Recall the inclusion
$\iota:V\hookrightarrow V_{\subscriptsizeBbb C}$ and the
${\Bbb C}^{\ast}$-action on $V_{\subscriptsizeBbb C}$ by multiplication.
Let $V=V_-\oplus V_+$ be an orthogonal decomposition of $V$ with
$\rho|_{V_-}$ negative-definite and $\rho|_{V_+}$ positive-definite.
Let $(e_1,\ldots,e_n)$ be a corresponding orthonormal basis for $V$
with, say, the first $p$ of them in $V_-$ and
the rest in $V_+$. Let
$$
 V_1\; =\; e^{-i\frac{3\pi}{4}}\cdot V_-\,
          +\, e^{-i\frac{\pi}{4}}\cdot V_+ \;
       =\; \mbox{\it Span}\,_{\subscriptsizeBbb R}(u_1,\ldots,u_n)
$$
and
$$
 V_2\; =\; i\cdot V_1\;
         =\; \mbox{\it Span}\,_{\subscriptsizeBbb R}(
                               u_1^{\prime},\ldots,u_n^{\prime})\,,
$$
where $u_j=e^{-i\frac{3\pi}{4}}e_j$ (resp.\ $e^{-i\frac{\pi}{4}}e_j$)
if $e_j\in V_-$ (resp.\ $V_+$) and $u_j^{\prime}=i\,u_j$.
Observe that $\rho_{\subscriptsizeBbb C}|_{V_1}$ is purely imaginary
negative-definite while $\rho_{\subscriptsizeBbb C}|_{V_2}$ is purely
imaginary positive-definite.

Let $v=(v_1,v_2)$ with respect to the decomposition
$V_{\subscriptsizeBbb C}=V_1\oplus V_2$. Then, with respect to the basis
$(u_1,\ldots,u_n;u_1^{\prime},\ldots,u_n^{\prime})$,
$$
 \rho_{\subscriptsizeBbb C}(v,v)\;
  =\; 2 v_1^t v_2\,+\,i\,(v_2^t v_2 - v_1^t v_1)\,.
$$
Consequently, if one defines $Q(v)=\rho_{\subscriptsizeBbb C}(v,v)$ and
lets $\overline{\rho}$ be the positive-definite inner product on
$V_{\subscriptsizeBbb C}$ that takes
$(u_1,\ldots,u_n;u_1^{\prime},\ldots,u_n^{\prime})$ as an orthonormal
frame, then
$$
 Q^{-1}({\Bbb R})\;
 =\; \left\{\,(v_1,v_2)\in V_{\subscriptsizeBbb C}\,|\,
  \overline{\rho}(v_1,v_1) = \overline{\rho}(v_2,v_2)\,\right\}\,.
$$
A real $n$-dimensional subspace $E$ in $V_{\subscriptsizeBbb C}$ has
the restriction $\rho_{\subscriptsizeBbb C}|_E$ real-valued if and only
if $E$ is contained in $Q^{-1}({\Bbb R})$. Such $E$ is characterized by
the following lemma.

\bigskip

\noindent
{\bf Lemma 2.1.} {\it $E \subset Q^{-1}({\Bbb R})$ if and only if
 $E$ is realizable as the graph of an isometry from
 $(V_1,\overline{\rho})$ to $(V_2,\overline{\rho})$.
}

\bigskip

\noindent
{\it Proof.} From the expression of $\rho_{\subscriptsizeBbb C}(v,v)$
 in terms of $(v_1,v_2)$, the if-part is clear. For the only-if part,
 observe that if $v=(v_1,v_2)$ is in $Q^{-1}({\Bbb R})$, then
 $\overline{\rho}(v_1,v_1)=\overline{\rho}(v_2,v_2)$. Consequently,
 if $v$ in $Q^{-1}({\Bbb R})$ is non-zero, then both $v_1$ and $v_2$
 must be non-zero. This implies that both the projections
 $\pr_1:E\rightarrow V_1$ and $\pr_2:E\rightarrow V_2$ induced from
 those of $V_{\subscriptsizeBbb C}$ into $V_1$ and $V_2$ are
 isomorphisms. The map $\pr_2\circ\pr_1^{-1}$ from
 $(V_1,\overline{\rho})$ to $(V_2,\overline{\rho})$ is then an isometry
 whose graph is $E$. This completes the proof.

\noindent\hspace{14cm}$\Box$

\bigskip

Since $\overline{\rho}$ is positive-definite, one has

\bigskip

\noindent
{\bf Corollary 2.2.} {\it The set $W$ of landing for Wick rotations
 is a natural $\mbox{\it GL}\,(n,{\Bbb R})$-bundle over $O(n)$, the real
 orthogonal group in dimension $n$.
}

\bigskip

We shall say more about the quadratic function $Q$ in Sec.\ 4. For now
let us study some details of $\overline{W}$.

\bigskip

\begin{flushleft}
{\bf A natural stratification of $\overline{W}$.}
\end{flushleft}
The space $\overline{W}$ can be decomposed into a collection of strata.
Each stratum contains all $E$ in $\overline{W}$ whose inner product
$\rho_{\subscriptsizeBbb C}$ is of a fixed type. With respect to the
basis $(u_1,\ldots,u_n)$ for $V_1$ and
$(u_1^{\prime},\ldots,u_n^{\prime})$ for $V_2$, the isometry from
$V_1$ to $V_2$ whose graph gives $E$ can be represented by an element
$A\in O(n)$. Furthermore if letting $x\in{\Bbb R}^n$ be the coordinates
on $E$ with respect to the basis $(\pr_1^{-1}u_j)_j$ (recall the proof
of Lemma 2.1), one has
$$
 Q(v)\; =\; 2\,x^t A x\; =\; x^t\, (A+A^t)\, x\,,
$$
for $v$ in $E$. In this way, the type $(r,s)$ of
$\rho_{\subscriptsizeBbb C}|_E$ corresponds exactly to the number of
positive and negative eigenvalues of the symmetrization
$(A+A^t)/\hspace{-.2ex}\mbox{\raisebox{-.3ex}{$2$}}$ of $A$. From this,
one can show that

\bigskip

\noindent{\bf Proposition 2.3.} {\it Let
 $\overline{W}_+=\mbox{\it SO}\,(n)$ and
 $\overline{W}_-=O_-(n)$ be the two components of $\overline{W}$. Let
 $\overline{W}_{r,s}$ be the subset in $\overline{W}$ that contains all
 $E$, on which $\rho_{\subscriptsizeBbb C}|_E$ is of type $(r,s)$. Then
 $\overline{W}_{r,s}$ is path-connected except
 $\overline{W}_{0,0}$, which has two components. Together they
 form a stratification of $\overline{W}$ with the following stratum
 relations, where $A\leftarrow B$ means that $B$ is contained in the
 closure of $A$.
 \begin{tabbing}
  \hspace{1em} \= $n$ {\rm :} odd  \\[1ex]
   \> \hspace{1em} $\mbox{\it SO}\,(n)$ {\rm :} \hspace{2em}
      \= $\overline{W}_{n,0}$ \= $\leftarrow$
                                             \= $\overline{W}_{n-2,0}$
      \= $\leftarrow$ \= $\overline{W}_{n-4,0}$ \= $\leftarrow$
      \= \hspace{1ex} $\cdots$ \hspace{1ex}\= $\leftarrow$
      \= $\overline{W}_{1,0}$      \\
   \> \> \> \> \hspace{1em} $\downarrow$ \>
      \> \hspace{1em} $\downarrow$ \> \> \hspace{.9em} $\downarrow$
      \> \> \hspace{.7em} $\downarrow$        \\
   \> \> \> \> $\overline{W}_{n-2,2}$ \> $\leftarrow$
                                            \> $\overline{W}_{n-4,2}$
      \> $\leftarrow$ \> \hspace{1ex} $\cdots$ \hspace{1ex}
      \> $\leftarrow$ \> $\overline{W}_{1,2}$ \\
   \> \> \> \> \> \> \hspace{1em} $\downarrow$ \>
      \> \hspace{.9em} $\downarrow$ \> \> \hspace{.7em} $\downarrow$ \\
   \> \> \> \> \> \> $\overline{W}_{n-4,4}$ \> $\leftarrow$
      \> \hspace{1ex} $\cdots$ \hspace{1ex} \> $\leftarrow$
      \> $\overline{W}_{1,4}$         \\
   \> \> \> \> \> \> \> \> \hspace{1ex} $\cdots$ \hspace{1ex} \>
      \> \hspace{1ex} $\cdots$      \\
   \> \> \> \> \> \> \> \> \hspace{.9em} $\downarrow$ \>
      \> \hspace{.7em} $\downarrow$      \\
   \> \> \> \> \> \> \> \> \hspace{-2.2ex} $\overline{W}_{3,n-3}$
      \> $\leftarrow$ \> $\overline{W}_{1,n-3}$      \\
   \> \> \> \> \> \> \> \> \> \> \hspace{.7em} $\downarrow$\\
   \> \> \> \> \> \> \> \> \> \> $\overline{W}_{1,n-1}$ ,\\[2em]
   \> \hspace{1em} $O_-(n)$ {\rm :} \hspace{2em}
      \> \hspace{-1.2em} $\overline{W}_{n-1,1}$ \> $\leftarrow$
      \> $\overline{W}_{n-3,1}$ \> $\leftarrow$
                                        \> $\overline{W}_{n-5,1}$
      \> $\leftarrow$ \> \hspace{1ex} $\cdots$ \hspace{1ex}
      \> $\leftarrow$ \> $\overline{W}_{0,1}$ \\
   \> \> \> \> \hspace{1em} $\downarrow$ \> \> \hspace{1em} $\downarrow$
      \> \> \hspace{.9em} $\downarrow$ \>
      \> \hspace{.7em} $\downarrow$\\
   \> \> \> \> $\overline{W}_{n-3,3}$ \> $\leftarrow$
                                             \> $\overline{W}_{n-5,3}$
      \> $\leftarrow$ \> \hspace{1ex} $\cdots$ \hspace{1ex}
      \> $\leftarrow$ \> $\overline{W}_{0,3}$ \\
   \> \> \> \> \> \> \hspace{1em} $\downarrow$ \>
      \> \hspace{.9em} $\downarrow$ \> \> \hspace{.7em} $\downarrow$ \\
   \> \> \> \> \> \> $\overline{W}_{n-5,5}$ \> $\leftarrow$
      \> \hspace{1ex} $\cdots$ \hspace{1ex} \> $\leftarrow$
      \> $\overline{W}_{0,5}$      \\
   \> \> \> \> \> \> \> \> \hspace{1ex} $\cdots$ \hspace{1ex} \>
      \> \hspace{1ex} $\cdots$      \\
   \> \> \> \> \> \> \> \> \hspace{.9em} $\downarrow$ \>
      \> \hspace{.7em} $\downarrow$    \\
   \> \> \> \> \> \> \> \> \hspace{-2.2ex} $\overline{W}_{2,n-2}$
      \> $\leftarrow$ \> $\overline{W}_{0,n-1}$      \\
   \> \> \> \> \> \> \> \> \> \> \hspace{.7em} $\downarrow$      \\
   \> \> \> \> \> \> \> \> \> \> $\overline{W}_{0,n}$ .
 \end{tabbing}

 \medskip 

 \begin{tabbing}
   \hspace{1em} \= $n$ {\rm :} even\\[1ex]
   \> \hspace{1em} $\mbox{\it SO}\,(n)$ {\rm :} \hspace{2em}
      \= $\overline{W}_{n,0}$ \= $\leftarrow$ \= $\overline{W}_{n-2,0}$
      \= $\leftarrow$ \= $\overline{W}_{n-4,0}$ \= $\leftarrow$
      \= \hspace{1ex} $\cdots$ \hspace{1ex}
      \= $\leftarrow$ \= $\overline{W}_{0,0}$  \\
   \> \> \> \> \hspace{1em} $\downarrow$ \>
      \> \hspace{1em} $\downarrow$ \> \> \hspace{.9em} $\downarrow$ \>
      \> \hspace{.7em} $\downarrow$  \\
   \> \> \> \> $\overline{W}_{n-2,2}$ \> $\leftarrow$
                                            \> $\overline{W}_{n-4,2}$
      \> $\leftarrow$ \> \hspace{1ex} $\cdots$ \hspace{1ex}
      \> $\leftarrow$ \> $\overline{W}_{0,2}$  \\
   \> \> \> \> \> \> \hspace{1em} $\downarrow$ \>
      \> \hspace{.9em} $\downarrow$ \> \> \hspace{.7em} $\downarrow$ \\
   \> \> \> \> \> \> $\overline{W}_{n-4,4}$ \> $\leftarrow$
      \> \hspace{1ex} $\cdots$ \hspace{1ex} \> $\leftarrow$
      \> $\overline{W}_{0,4}$    \\
   \> \> \> \> \> \> \> \> \hspace{1ex} $\cdots$ \hspace{1ex} \>
      \> \hspace{1.2ex} $\cdots$   \\
   \> \> \> \> \> \> \> \> \hspace{.9em} $\downarrow$ \>
      \> \hspace{.7em} $\downarrow$    \\
   \> \> \> \> \> \> \> \> \hspace{-1ex}$\overline{W}_{2,n-2}$
      \> $\leftarrow$ \> $\overline{W}_{0,n-2}$    \\
   \> \> \> \> \> \> \> \> \> \> \hspace{.7em} $\downarrow$    \\
   \> \> \> \> \> \> \> \> \> \> $\overline{W}_{0,n}$ , \\[2em]
   \> \hspace{1em} $O_-(n)$ {\rm :} \hspace{2em}
      \> \hspace{-1.2em} $\overline{W}_{n-1,1}$ \> $\leftarrow$
      \> $\overline{W}_{n-3,1}$ \> $\leftarrow$
                                             \> $\overline{W}_{n-5,1}$
      \> $\leftarrow$ \> \hspace{1ex} $\cdots$ \hspace{1ex}
      \> $\leftarrow$ \> $\overline{W}_{1,1}$     \\
   \> \> \> \> \hspace{1em} $\downarrow$ \>
      \> \hspace{1em} $\downarrow$ \> \> \hspace{.9em} $\downarrow$ \>
      \> \hspace{.7em} $\downarrow$    \\
   \> \> \> \> $\overline{W}_{n-3,3}$ \> $\leftarrow$
                                             \> $\overline{W}_{n-5,3}$
      \> $\leftarrow$ \> \hspace{1ex} $\cdots$ \hspace{1ex}
      \> $\leftarrow$ \> $\overline{W}_{1,3}$ \\
   \> \> \> \> \> \> \hspace{1em} $\downarrow$ \>
      \> \hspace{.9em} $\downarrow$ \> \> \hspace{.7em} $\downarrow$ \\
   \> \> \> \> \> \> $\overline{W}_{n-5,5}$ \> $\leftarrow$
      \> \hspace{1ex} $\cdots$ \hspace{1ex} \> $\leftarrow$
      \> $\overline{W}_{1,5}$     \\
   \> \> \> \> \> \> \> \> \hspace{1ex} $\cdots$ \hspace{1ex} \>
      \> \hspace{1ex} $\cdots$     \\
   \> \> \> \> \> \> \> \> \hspace{.9em} $\downarrow$ \>
      \> \hspace{.7em} $\downarrow$   \\
   \> \> \> \> \> \> \> \> \hspace{-1.7ex} $\overline{W}_{3,n-3}$
      \> $\leftarrow$ \> $\overline{W}_{1,n-3}$      \\
   \> \> \> \> \> \> \> \> \> \> \hspace{.7em} $\downarrow$      \\
   \> \> \> \> \> \> \> \> \> \> $\overline{W}_{1,n-1}$ .
 \end{tabbing}
} 

\bigskip

\noindent
{\it Remark 2.4.} Notice that the multiplication by $i$ in
$V_{\subscriptsizeBbb C}$ induces a homeomorphism between
$\overline{W}_{r,s}$ and $\overline{W}_{s,r}$.

\bigskip

\noindent
{\it Proof.} Let us outline the idea of the proof first. The
$\mbox{\it SO}\,(n)$-action on $O(n)$ by conjugation commutes with
symmetrization; thus it leaves each $\overline{W}_{r,s}$
invariant. Since $\mbox{\it SO}\,(n)$ is path-connected, each orbit of
this action, i.e.\ a conjugacy class in $O(n)$, also has to be
path-connected. Consequently, the quotient map for this action,
$O(n)\rightarrow O(n)/\hspace{-.2ex}\mbox{\raisebox{-.3ex}{$\sim$}}$
pushes the decomposition $\{\overline{W}_{r,s}\}$ of $\overline{W}$
down to the decomposition
$\{\overline{W}(r,s)/\hspace{-.2ex}\mbox{\raisebox{-.3ex}{$\sim$}}\}$
of $\overline{W}/\hspace{-.2ex}\mbox{\raisebox{-.3ex}{$\sim$}}$ with
the same stratum relations; and the corresponding pieces
$\overline{W}_{r,s}$ and
$\overline{W}_{r,s}/\hspace{-.2ex}\mbox{\raisebox{-.3ex}{$\sim$}}$
have the same number of components. This reduces the problem to the
study of $O(n)/\hspace{-.2ex}\mbox{\raisebox{-.3ex}{$\sim$}}$ with the
quotient decomposition
$\{\overline{W}(r,s)/\hspace{-.2ex}\mbox{\raisebox{-.3ex}{$\sim$}}\}$.
To accomplish this, a key fact is that [Hu]
$$
 \mbox{\it SO}\,(n)/
   \hspace{-.2ex}\mbox{\raisebox{-.3ex}{$\sim$}}\;
 = \; \mbox{\raisebox{.3ex}{$T/$}}\mbox{\raisebox{-.3ex}{$\Gamma$}}\,,
$$
where $T$ is a maximal torus in $\mbox{\it SO}\,(n)$ and $\Gamma$ is
the Weyl group acting on $T$. Up to conjugations, elements in $T$ is
built up from elements in $U(1)$, represented by the $2\times 2$ blocks
$D(\theta)=\left(\begin{array}{rc}
             \cos\theta & \sin\theta \\ -\sin\theta & \cos\theta
           \end{array}  \right)$,
whose symmetrization is
$\left(\begin{array}{cc}
         \cos\theta &  0 \\ 0 & \cos\theta
       \end{array}  \right)$.
This gives a stratification of $U(1)$ by the sign of $\cos\theta$.
The stratification of
$\overline{W}/\hspace{-.2ex}\mbox{\raisebox{-.3ex}{$\sim$}}$ is induced
from the product of these stratified $U(1)$'s; hence so is that of
$\overline{W}$. This also shows that the jump of the type of
$\rho_{\subscriptsizeBbb C}|_E$ for $E$ in various strata of
$\overline{W}$ is a multiple of $2$. ({\sc Figure 2-1}.)

\begin{figure}[htbp]
\setcaption{{\sc Figure 2-1.}
\baselineskip 14pt 
  The fundamental stratification of
  $U(1)$ whose product leads to the stratification of $\overline{W}$.
  The influence to the type of metrics is also indicated.}
\centerline{\psfig{figure=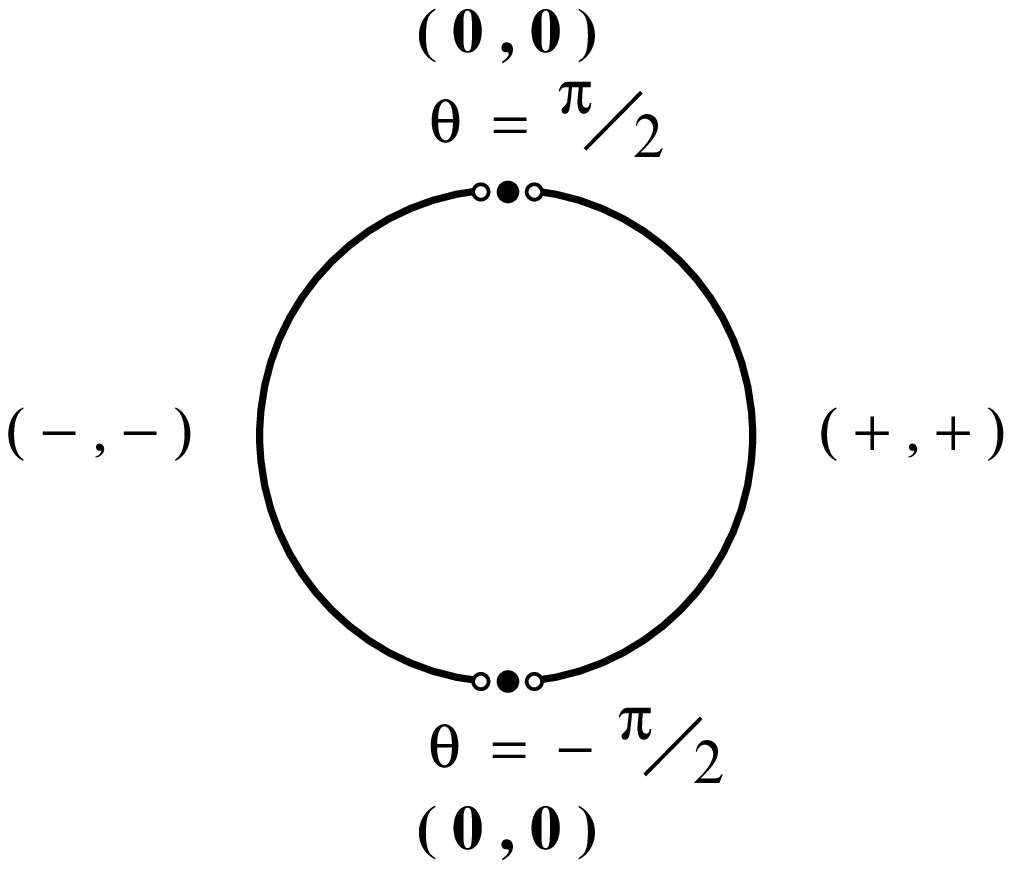,width=13cm,caption=}}
\end{figure}

Let us now look at the details.

\bigskip

\noindent
{\it Case (i): $n$ odd.}
For $n=2k+1$, we may choose $T$ consisting of the block diagonal
matrices:
\begin{eqnarray*}
  T & = & \{ \mbox{\it Diag}\,
   (1, D(\theta_1), \cdots, D(\theta_k))\vert\theta_i\in [0,2\pi] \}\\
     & \cong & \underbrace{S^1\times\cdots\times S^1}_{k}\,;
\end{eqnarray*}
then $\Gamma$ consists of the $2^kk!$ permutations of the indices of
$(\theta_1,\cdots,\theta_k)$ composed with substitutions
$$
 (\theta_1,\cdots,\theta_k)\; \longrightarrow\;
                (\epsilon_1\theta_1,\cdots,\epsilon_k\theta_k)\,,
$$
where $\epsilon_i=\pm 1$. It follows that the quotient
$\mbox{\it SO}\,(2k+1)/\hspace{-.2ex}\mbox{\raisebox{-.3ex}{$\sim$}}$
is stratified by the collection of subsets $I_{\alpha\beta}$ defined
by
\begin{eqnarray*}
 \lefteqn{ I_{\alpha\beta}\; =\; \{(\theta_1,\cdots,\theta_k)\vert 0
  \leq\theta_1\leq\cdots\leq\theta_{\alpha} }\\
 & & \hspace{2cm}
        <\frac{\pi}{2} =\theta_{\alpha+1}=\cdots=\theta_{k-\beta}
               <\theta_{k-\beta+1}\leq\cdots\leq\theta_k\leq\pi\}\,.
\end{eqnarray*}
One can check that $I_{\alpha\beta}$ is path-connected and lies in
$\overline{W}_{2\alpha+1,2\beta}/
                   \hspace{-.2ex}\mbox{\raisebox{-.3ex}{$\sim$}}$.
From the symmetry $(r,s)\leftrightarrow (s,r)$, one also obtains
the stratum structure in question for $O_-(2k+1)$.
By comparing the even- or oddness of $r$ and $s$, one can see that
$\overline{W}(r,s)/\hspace{-.2ex}\mbox{\raisebox{-.3ex}{$\sim$}}$
lies either in $\mbox{\it SO}\,(n)$ or $O_-(n)$; thus it must be that
$$
 I_{\alpha\beta}\;
  =\; \overline{W}_{2\alpha+1,2\beta}/
                  \hspace{-.2ex}\mbox{\raisebox{-.3ex}{$\sim$}}\,.
$$
This shows that
$\overline{W}_{2\alpha+1,2\beta}/
            \hspace{-.2ex}\mbox{\raisebox{-.3ex}{$\sim$}}$
is path-connected. One easily sees that the relations among these
strata, which are the same as those among $I_{\alpha\beta}$, are as
indicated in the Proposition. This completes the proof for $n$ odd.

\bigskip

\noindent
{\it Case (ii): $n$ even.}
For $n=2k$, consider first the $\mbox{\it SO}\,(2k)$-part, we may choose
$$
 T \;=\; \{ \mbox{\it Diag}\,(D(\theta_1), \cdots, D(\theta_k))\vert
                                     \theta_i\in [0,2\pi] \}\,,
$$
then $\Gamma$ consists of the $2^{k-1}k!$ permutations of the indices
of $(\theta_1,\cdots,\theta_k)$ composed with substitutions
$$
 (\theta_1,\cdots,\theta_k)\; \longrightarrow\;
                  (\epsilon_1\theta_1,\cdots,\epsilon_k\theta_k)
$$
with $\epsilon_i = \pm1$ and $\epsilon_1\cdots\epsilon_k = 1$. From
these data, it follows that
\begin{eqnarray*}
  \mbox{\raisebox{.3ex}{$T/$}}\mbox{\raisebox{-.3ex}{$\Gamma$}}
 & \cong  & \Delta_1\cup_h\Delta_2,\quad\mbox{where}  \\
 \Delta_1  & =  & \{(\theta_1,\cdots,\theta_k)\vert
       0\leq\theta_1\leq\cdots \leq\theta_k\leq\pi\}\,,  \\
 \Delta_2  & =  & \{(\theta_1,\cdots,\theta_k)\vert
         \pi\leq\theta_k\leq 2\pi,\;
        0\leq\theta_1\leq\cdots\leq\theta_{k-1}\leq 2\pi-\theta_k\}
\end{eqnarray*}
and $h$ is the pasting map from
$\{\theta_1=0\}$-face$\,\cup\,$$\{\theta_k=\pi\}$-face of $\Delta_1$ to
that of $\Delta_2$ defined by
\begin{eqnarray*}
 h(0,\theta_2,\cdots,\theta_{k-1},\theta_k) & =
    & (0,\theta_2,\cdots,\theta_{k-1},2\pi-\theta_k) \quad\mbox{and}\\
 h(\theta_1,\cdots,\theta_{k-1},\pi)  & =
               & (\theta_1,\cdots,\theta_{k-1},\pi)\,.
\end{eqnarray*}
The two collections of subsets: $I_{\alpha\beta}$ in $\Delta_1$ and
$I_{\alpha\beta}^{\prime}$ in $\Delta_2$ defined by
\begin{eqnarray*}
 \lefteqn{I_{\alpha\beta}\; =\; \{(\theta_1,\cdots,\theta_k)\vert
                      0 \leq\theta_1\leq\cdots\leq\theta_{\alpha} }\\
  & & \hspace{3cm}<\frac{\pi}{2} = \theta_{\alpha+1} = \cdots
  = \theta_{k-\beta}<\theta_{k-\beta+1}\leq\cdots\leq\theta_k\leq\pi\}
\end{eqnarray*}
and
$$
 I^{\prime}_{\alpha\beta}\; =\; \mbox{the reflection of}\,\,
     I_{\alpha\beta} \,\,\mbox{with respect to the hyperplane}\,\,
                 \theta_k=\pi\,,
$$
form a stratum structure for $T/\mbox{\raisebox{-.3ex}{$\Gamma$}}$
after being pasted along the map $h$. By the same reason as in the case
of $n$ odd, $I_{\alpha\beta}$ and $I^{\prime}_{\alpha\beta}$ together
constitute
$\overline{W}_{2\alpha,2\beta}/
       \hspace{-.2ex}\mbox{\raisebox{-.3ex}{$\sim$}}$.
Thus
$\overline{W}_{2\alpha,2\beta}/
                      \hspace{-.2ex}\mbox{\raisebox{-.3ex}{$\sim$}}$
is disconnected if and only if both $I_{\alpha\beta}$ and
$I^{\prime}_{\alpha\beta}$ have empty intersection with
$\{\theta_1=0\}$ and $\{\theta_k=\pi\}$. This implies that
($\frac{\pi}{2}\leq\theta_1\leq\pi$) and
($0\leq\theta_k\leq\frac{\pi}{2}$ or
$\frac{3\pi}{2}\leq\theta_k\leq 2\pi$). Together with the defining
inequalities for $\Delta_1$ and $\Delta_2$, we have either
$\theta_1=\theta_2=\cdots=\theta_k=\frac{\pi}{2}$ or
$\theta_1=\theta_2=\cdots=\theta_{k-1}=\frac{\pi}{2},
 \,\theta_k=\frac{3\pi}{2}$, which correspond to $I_{00}$ and
$I^{\prime}_{00}$ respectively.
Thus
$\overline{W}_{r,s}/\hspace{-.2ex}\mbox{\raisebox{-.3ex}{$\sim$}}$
is path-connected except
$\overline{W}_{0,0}/\hspace{-.2ex}\mbox{\raisebox{-.3ex}{$\sim$}}$,
which has two components.

For the $O_-(2k)$ part, any $A$ in $O_-(2k)$ is conjugate under
$\mbox{\it SO}\,(2k)$ to some $\mbox{\it Diag}\,(-1,B)$, where
$B\in \mbox{\it SO}\,(2k-1)$. Thus
$O_-(2k)/\hspace{-.2ex}\mbox{\raisebox{-.3ex}{$\sim$}}$
is actually a quotient of
$\mbox{\raisebox{.3ex}{$T^{\prime}/$}}
                      \mbox{\raisebox{-.3ex}{$\Gamma^{\prime}$}}$,
where $T^{\prime}$ is a maximal torus of $\mbox{\it SO}\,(2k-1)$ and
$\Gamma^{\prime}$ is the Weyl group acting on $T^{\prime}$. The
stratum structure descends then from that for $n$ odd. As already shown
in the odd case, each stratum in
$\mbox{\raisebox{.3ex}{$T^{\prime}/$}}
                    \mbox{\raisebox{-.3ex}{$\Gamma^{\prime}$}}$
is path-connected; hence so is its quotient. This implies that all the
corresponding
$\overline{W}_{r,s}/\hspace{-.2ex}\mbox{\raisebox{-.3ex}{$\sim$}}$
in $O_-(2k)$ are path-connected. The stratum relations also follow.
This completes the case of $n$ even and hence the proof of the
proposition.

\noindent\hspace{14cm}$\Box$

\bigskip

\noindent
{\it Remark 2.5.} For a manifold $M=(M,\rho)$, one can apply the above
discussion fiberwise and construct the {\it bundle of landing $W(M)$}
and the $O(n)$-bundle $\overline{W}(M)$ over $M$. The latter has the
structure group $O(n)$, which acts on the $O(n)$-fiber by conjugation.
Consequently, $\overline{W}(M)$ has two components
$\overline{W}_+(M)$, whose fiber is $\mbox{\it SO}\,(n)$, and
$\overline{W}_-(M)$, whose fiber is $O_-(n)$. The stratification of
$\overline{W}$ is invariant under conjugation; and hence it leads a
stratification of $\overline{W}(M)$ by subbundles
$\overline{W}_{r,s}(M)$ whose fiber is $\overline{W}_{r,s}$.

\bigskip

So far our attention has been mainly to the set of landing $W$ itself.
However a complete geometric picture for Wick rotations involves also
how $W$ embeds in $M(n,{\Bbb C})$ or $\mbox{\it GL}\,(n,{\Bbb C})$. This
should be a problem that falls well in the realm of the study of
singularities of hypersurface intersections and their complement in
${\Bbb C}^n$. In Sec.\ 4, we shall study this problem in dimension two.

\bigskip

\section{$\overline{W}$ in two, three, and four dimensions.}

As corollaries and supplements to the previous section, let us provide
concrete calculations of $\overline{W}$ at two, three, and four
dimensions. We shall denote $\overline{W}$ for $V$ of dimension $n$ by
$\overline{W}(n)$ and its two components by $\overline{W}_+(n)$ - the
$\mbox{\it SO}\,(n)$-part - and $\overline{W}_-(n)$ - the $O_-(n)$-part.

\bigskip

\begin{flushleft}
{\bf In two dimensions.}
\end{flushleft}
It is clear that the topology and stratification of $\overline{W}(2)$
is given by
$$
\begin{array}{rccccc}
 \overline{W}_+(2)\,: & & \overline{W}_{2,0}\cong\stackrel{\circ}{I}
                  & \leftarrow &\overline{W}_{0,0}\cong\partial I\\
    & & & & \downarrow &\\
    & & & & \overline{W}_{0,2}\cong\stackrel{\circ}{I}  &\\
    & & & & & \\
 \overline{W}_-(2)\,: & & \overline{W}_{1,1}\cong S^1  & & &,
\end{array}
$$
where $\stackrel{\circ}{I}$ is an open interval ({\sc Figure 3-1}).

\begin{figure}[htbp]
\setcaption{{\sc Figure 3-1.}
\baselineskip 14pt  
   The stratification of $\overline{W}(2)$
   and the corresponding types of inner products.     }
\centerline{\psfig{figure=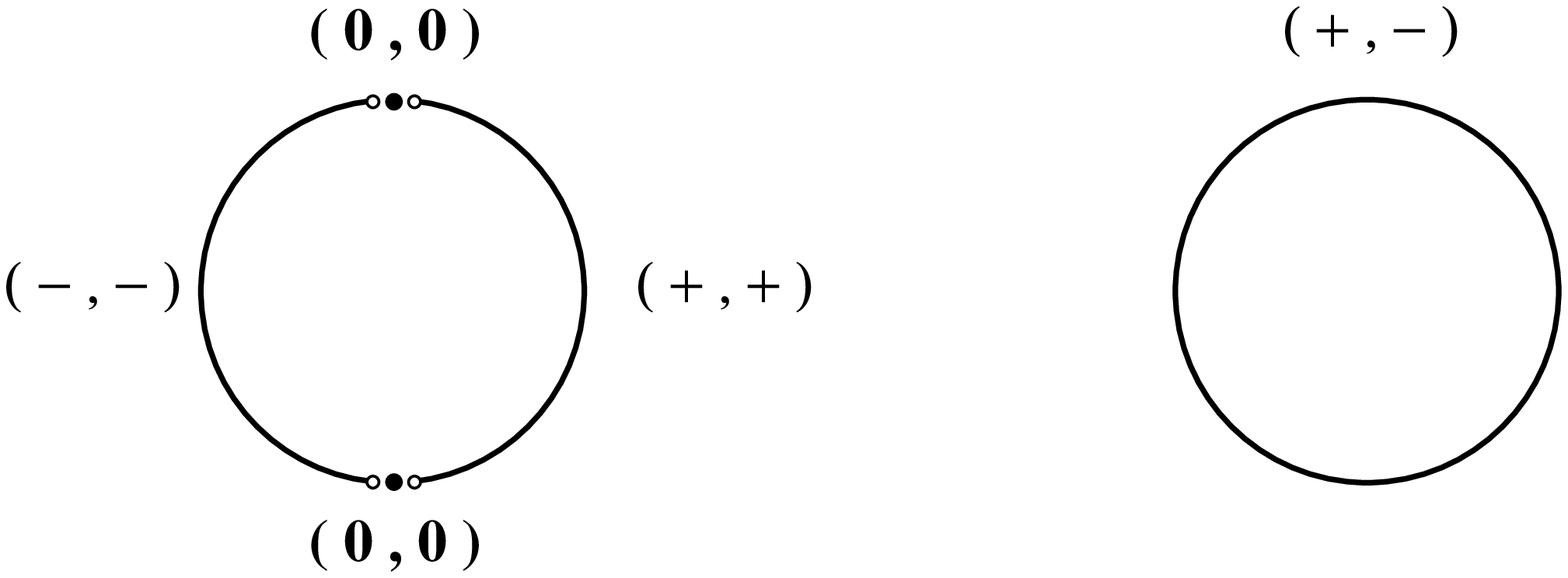,width=13cm,caption=}}
\end{figure}

\bigskip

\begin{flushleft}
{\bf In three dimensions.}
\end{flushleft}
Recall that $\mbox{\it SO}\,(3)=\mbox{{\Bbb R}P}^3$ has a model as the
closed 3-ball $D^3$
of radius $\pi$ with the antipodal points on the boundary $\partial D^3$
identified. With respect to the coordinates $({\bf n},\theta)$,
where $\theta\in [0,\pi]$ and ${\bf n} \in S^2$, $({\bf n},\theta)$
represents the rotation along the axis ${\bf n}$ by an angle $\theta$
following the right-hand rule. It is conjugate to
$\mbox{\it Diag}\,(1,D(\theta))$. For $O_-(3)$, the same model still
works with small modifications: the ${\bf n}$ is now the
$(-1)$-eigenvector of an element in $O_-(3)$ and $\theta=\pi$, which
corresponds to $-\mbox{Identity}$, is now at the origin of the ball.
Thus $({\bf n},\theta)$ is conjugate to
$\mbox{\it Diag}\,(-1,D(\theta))$. From these,
$\overline{W}_+(3)$ with its stratification is given by
$$
\begin{array}{lcrcll}
 \overline{W}_+(3):  &
    & \overline{W}_{3,0}\cong\stackrel{\circ}{D}_{\frac{\pi}{2}}
    & \leftarrow   & \overline{W}_{1,0}\cong S^2  &  \\
 & & & & \hspace{4em}\downarrow   & \\
 & & &
     & \overline{W}_{1,2} \cong
       \mbox{{\Bbb R}P}^3 - D_{\frac{\pi}{2}} & ,\\
\end{array}
$$
where $D_{\frac{\pi}{2}}$ is the closed 3-ball of radius $\frac{\pi}{2}$
in our model and $\stackrel{\circ}{D}_{\frac{\pi}{2}}$ is its interior.
And the $\overline{W}_-(3)$ part can be obtained by the symmetry
$(r,s)\leftrightarrow (s,r)$. ({\sc Figure 3-2}.)

\begin{figure}[htbp]
\setcaption{{\sc Figure 3-2.}
\baselineskip 14pt  
   The stratification of $\overline{W}_+(3)$
   and the various corresponding types of inner product. In the picture,
   $\overline{W}_+(3)={\Bbb R}{\rm R}^3$ is realized as a $3$-ball with
   each pair of antipodal points on the boundary identified. The 
   $\overline{W}_-(3)$-part is similar with $\pm$ changed to $\mp$.  }
\centerline{\psfig{figure=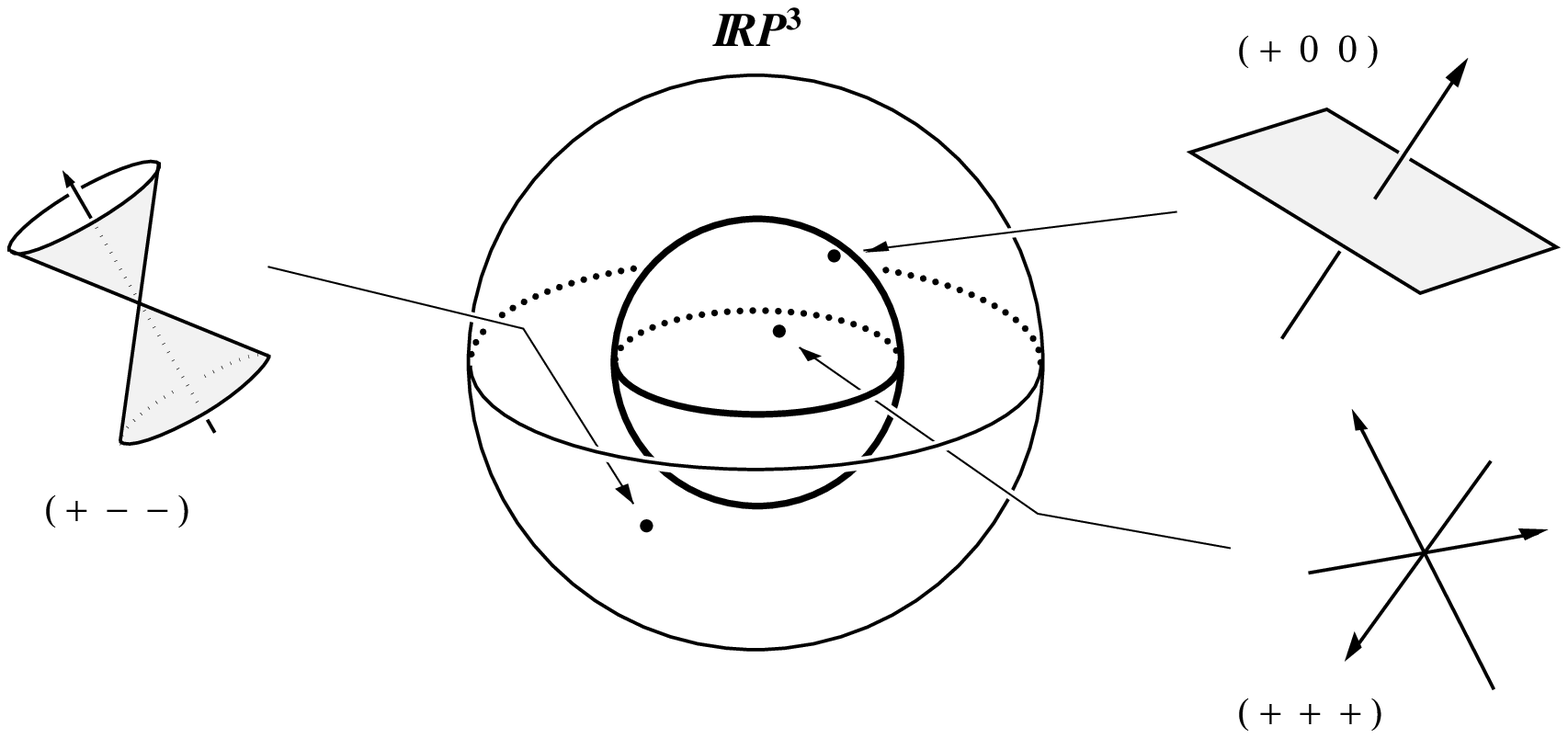,width=13cm,caption=}}
\end{figure}

\bigskip

\begin{flushleft}
{\bf In four dimensions.}
\end{flushleft}
\noindent{\bf (a) The $\overline{W}_+(4)$ part.} Recall that
$\mbox{\it SO}\,(4)\cong S^3\times \mbox{\it SO}\,(3)$ can be
interpreted as the group of ${\Bbb R}$-linear, orientation and
norm-preserving self-maps on quaternions. In the above topological
product decomposition, the $S^3$-factor represents the unit quaternion
group, which acts on quaternions by left multiplication; the
$\mbox{\it SO}\,(3)$-factor represents the isotropy group of 1, which
fixes the real axis and rotates the ${\Bbb R}^3$-space spanned by $i$,
$j$, and $k$. In the following discussion, we shall not distinguish the
normed space of quaternions and the Euclidean 4-space ${\Bbb E}^4$ with
$(1,i,j,k)$ as an orthonormal basis.

Consider the coset $S^3\cdot h$ in $\mbox{\it SO}\,(4)$ with
$h\in \mbox{\it SO}\,(3)$. Represent $h$ by $({\bf n},\theta)$ as
introduced in the discussion for $\overline{W}(3)$. Given
$q\in S^3$, let
\begin{quote}
 \hspace{-.5em}
 (1) \hspace{1em} \parbox[t]{13cm}{
  $a=\mbox{\rm Re}\,q$, the real part of $q$; }

 \hspace{-.5em}
 (2) \hspace{1em} \parbox[t]{13cm}{
  $b=\langle q,{\bf n}\rangle$, where both $q$ and ${\bf n}$ are
  regarded as vectors in ${\Bbb E}^4$; and}

 \hspace{-.5em}
 (3) \hspace{1em} \parbox[t]{13cm}{
  $A$ be the matrix in $\mbox{\it SO}\,(4)$ that represents $qh$. }
\end{quote}
Notice that $a^2+b^2\leq 1$. It follows from straightforward
computation that $\frac{A+A^t}{2}$ is conjugate to the diagonal matrix
$\mbox{\it Diag}\,(\cos\theta_1, \cos\theta_1, \cos\theta_2,
 \cos\theta_2)$
with
$$
 \cos\theta_1 \;=\; \frac{1}{2}\left(a+a\cos\theta-b\sin\theta+
     \sqrt{(a-a\cos\theta+b\sin\theta)^2
           +4(1-a^2-b^2)\sin^2\frac{\theta}{2}} \right)
$$
and
$$
 \cos\theta_2 \; =\; \frac{1}{2}\left(a+a\cos\theta-b\sin\theta
    - \sqrt{(a-a\cos\theta+b\sin\theta)^2
                  +4(1-a^2-b^2)\sin^2\frac{\theta}{2}}  \right).
$$
Using this, consider the following maps:
$$
\begin{array}{cccccl}
 S^3  & \stackrel{\eta_{({\bf n},\theta)}}{\longrightarrow}  & D^2
          & \stackrel{\zeta}{\longrightarrow}   & \Delta  & \\[1ex]
  q   & \longmapsto   & (a,b)   & \longmapsto
                                  & (\cos\theta_1,\cos\theta_2) &,
\end{array}
$$
where $D^2$ is the unit disk and
$\Delta=\{(s,t)\vert s,t \in [-1,1], s\geq t\}$.
For $\theta\in (0,\pi)$, one has in $D^2$
\begin{eqnarray*}
 \{\cos\theta_1 = 0\}  & =
        & \{ a\cos\frac{\theta}{2} - b\sin\frac{\theta}{2}
                    = -\sin\frac{\theta}{2}\,,\quad a^2+b^2\leq 1\}\,,
\\ \{\cos\theta_2 = 0\}  & =
        & \{ a\cos\frac{\theta}{2} - b\sin\frac{\theta}{2}
                    =\sin\frac{\theta}{2}\,,\quad a^2+b^2\leq 1\}\,.
\end{eqnarray*}
For $\theta=0$, one has
$$
 \{\cos\theta_1 = 0\} \; =\;  \{\cos\theta_2 = 0\} \;
     =\;  \{ a=0\,,\; b\leq 1 \} \,.
$$
For $\theta=\pi$, one has
$$
 \{\cos\theta_1 = 0\} \; =\;  \{\cos\theta_2 = 0\} \;
                                    =\; \{ (0,-1), (0,1) \}\,.
$$
Decomposing $D^2$ by
$(\mbox{\it sign}\,(\cos\theta_1),\mbox{\it sign}\,(\cos\theta_2))$
under the map $\zeta$ and considering the preimage under
$\eta_{({\bf n},\theta)}$ of the components of the decomposition, one
then has:

\begin{quote}
 \hspace{-1.5em}(i) $\overline{W}_{4,0}\cong \overline{W}_{0,4}\cong\;$
  a $\stackrel{\circ}{D^3}$ bundle over $\stackrel{\circ}{D^3}$,
  which must be
  $\stackrel{\circ}{D^3}\times\stackrel{\circ}{D^3}
   \cong \stackrel{\circ}{D^6}$. The base manifold
  $\stackrel{\circ}{D^3}$ is $\mbox{{\Bbb R}P}^3-\{\theta=\pi\}$.

 \hspace{-1.6em}(ii) $\overline{W}_{2,2}\cong\;$ a two-punctured $S^3$
  bundle over a one-punctured $\mbox{{\Bbb R}P}^3$. For
  $\theta\ne 0,\pi$, the two punctures on each fiber correspond to two
  sections in the trivial $S^3$ bundle,
  $(\mbox{{\Bbb R}P}^3-\{\theta=0,\pi\})\times S^3$: one comes from
  $$
  \begin{array}{ccccc}
    \sigma_1   & :   & \mbox{{\Bbb R}P}^3-\{\theta=0,\pi\}
                                    & \longrightarrow   & S^3  \\
    & & ({\bf n},\theta)   & \longmapsto
    & \eta^{-1}_{({\bf n},\theta)}(\cos\frac{\theta}{2},
                                            -\sin\frac{\theta}{2})
  \end{array}
  $$
  and the other comes from
  $$
  \begin{array}{ccccc}
   \sigma_2   & :   & \mbox{{\Bbb R}P}^3-\{\theta=0,\pi\}
                                   & \longrightarrow    & S^3  \\
   & & ({\bf n},\theta)   & \longmapsto
     &\eta^{-1}_{({\bf n},\theta)}(-\cos\frac{\theta}{2},
                                             \sin\frac{\theta}{2})\;.
  \end{array}
  $$
  Since $({\bf n},\pi)=(-{\bf n},\pi)$,
  $$
   \lim_{\theta\to\pi}\sigma_1({\bf n},\theta) \;
     =\;     -\lim_{\theta\to\pi}\sigma_2({\bf n},\theta)\,.
  $$
  Thus $\sigma_1$, $\sigma_2$ together with their extension
  over $\{\theta=\pi\}$ form a connected double covering $X$ over
  $\mbox{{\Bbb R}P}^3-\{\theta=0\}$. Thus $X\cong S^2\times (-1,1)$
  and $\overline{W}_{2,2}\cong$ the complement bundle of $X$ in
  $(\mbox{{\Bbb R}P}^3-\{\theta=0\})\times S^3$.

  \hspace{-1.8em}(iii)
  $\overline{W}_{2,0}\cong \overline{W}_{0,2} \cong\;$ the bundle over
  $\mbox{{\Bbb R}P}^3-\{\theta=0,\pi\}$ with fibre the boundary of the
  fibre of the trivial bundle $\overline{W}_{4,0}$; hence it is
  homeomorphic to $(\stackrel{\circ}{D^3}-\{\theta=0\})\times S^2$.

  \hspace{-1.8em}(iv) $\overline{W}_{0,0}$ lies over
  $\{\theta=0\}\cup\{\theta=\pi\}$. Over $\theta=0$, it is the
  preimage of $\{a=0, \vert b\vert\leq 1\}$ under
  $\eta_{({\bf n}_0,0)}$, where ${\bf n}_0$ is any fixed
  direction. This is a 2-sphere. Over $\theta=\pi$, it is a connected
  double covering over $\mbox{{\Bbb R}P}^2$. Thus it is also a
  2-sphere. Consequently, $\overline{W}_{0,0}$ is the disjoint union
  $S^2\amalg S^2$.
\end{quote}

\bigskip

\noindent{\bf (b) The $\overline{W}_-(4)$ part.} Let $A=qh$ be a matrix in
$O_-(4)$ that lies in the coset $S^3\cdot h$, where $h\in O_-(3)$. With
the similar notation and argument as in the $\overline{W}_+(4)$ part
(${\bf n}$ is now the $(-1)$-eigenvector of $h$), $\frac{A+A^t}{2}$ is
conjugate to $\mbox{\it Diag}\,(-1,1,\cos\xi,\cos\xi)$, where
$\cos\xi = a\cos\theta - b\sin\theta$. Similar to the
$\overline{W}_+(4)$ part, one has maps
$$
 \begin{array}{cccccl}
  S^3   & \stackrel{\eta_{({\bf n},\theta)}}{\longrightarrow}
     & D^2   & \stackrel{\zeta}{\longrightarrow}   & [-1,1] & \\[1ex]
  q   & \longmapsto   & (a,b)   & \longmapsto   & \cos\xi   &.
 \end{array}
$$
and also the decomposition of $D^2$ by $\mbox{\it sign}\,(\cos\xi)$.
The preimage under $\eta_{({\bf n},\theta)}$ of the $(+)$-region
(resp.\ the $(-)$-region) is the hemisphere with center
$\eta^{-1}_{({\bf n},\theta)}(\cos\theta,-\sin\theta)$, (resp.\
$\eta^{-1}_{({\bf n},\theta)}(-\cos\theta,\sin\theta)$).

Let
$$
\begin{array}{cccccl}
 \sigma   & :   & \mbox{{\Bbb R}P}^3   & \longrightarrow   & S^3
                                                            &\\[1ex]
  & & ({\bf n},\theta)   & \longmapsto
          & \eta^{-1}_{({\bf n},\theta)}(\cos\theta,-\sin\theta) & .
\end{array}
$$
It is straightforward to check that $\sigma$ is well-defined and
is of degree $1$. Regard $\overline{W}_{3,1}$ as a subbundle in the
trivial bundle $\mbox{{\Bbb R}P}^3\times S^3$ over
$\mbox{{\Bbb R}P}^3$ with coordinates
$(x,y), x\in \mbox{{\Bbb R}P}^3, y\in S^3$. Recall the unit
quaternion group structure on $S^3$. The left multiplication then takes
a hemisphere to another hemisphere. Let $\Omega$ now be the open
hemisphere in $S^3$ centered at 1. The map
$$
\begin{array}{ccc}
 \overline{W}_{3,1}   & \longrightarrow
                              & \mbox{{\Bbb R}P}^3\times\Omega\\
 (x,y)   & \longmapsto   & (x,\sigma(x)^{-1}\cdot y)
\end{array}
$$
gives a bundle homeomorphism which trivializes $\overline{W}_{3,1}$.
Thus
$$
 \overline{W}_{3,1}\; \cong\; \overline{W}_{1,3}\;
    \cong\; \mbox{{\Bbb R}P}^3\times\stackrel{\circ}{D^3}\,.
$$
Followingly, one also has                       
$$
 \overline{W}_{1,1}\; \cong\; \mbox{{\Bbb R}P}^3\times S^2\,.
$$

This completes the discussion for $\overline{W}(4)$.

\bigskip

\section{Wick rotations in two dimensions.}

In this last section, we shall study more the geometric objects associated
to Wick rotations in two dimensions. Following the notations from Sec.\ 1
and 2, we recall from Sec.\ 2 and 3 that every $(\xi_1,\xi_2)$ in $W(2)$
lies in a real $2$-subspace $E$ in $V_{\subscriptsizeBbb C}$ that is
realizable as the graph of an isometry $\phi$ from $(V_1,\overline{\rho})$
to $(V_2,\overline{\rho})$. In two dimensions, two different such $E_1$ and
$E_2$ can have non-trivial intersection only if one is in
$\overline{W}_+(2)=\mbox{\it SO}\,(2)$ and the other in
$\overline{W}_-(2)=O_-(2)$. The two points in $\overline{W}_+(2)$ labelled
by $(0,0)$ in {\sc Figure 3-1} correspond to the two $E$'s in
$\overline{W}$ that happen to be a complex line. From Corollary 2.2, $W(2)$
with all the $\Bbb R$-linearly dependent $(\xi_1,\xi_2)$ removed is a
trivial $\mbox{\it GL}\,(2,{\Bbb R})$-bundle over $\overline{W}(2)$.
Let $W_+(2)$ be the part over $\overline{W}_+(2)$ and $W_-(2)$ be the part
over $\overline{W}_-(2)$. Then, from Sec.\ 3, we know that $W(2)-\Xi$ has
six components, four for $W_+(2)-\Xi$ and two for $W_-(2)-\Xi$. However, all
together, $W(2)$ is a connected subset in $M(2,{\Bbb C})$. Indeed, since
any non-zero $v\in Q^{-1}({\Bbb R})$ determines exactly one
orientation-preserving and one orientation-reversing isometry from
$(V_1,\overline{\rho})$ to $(V_2,\overline{\rho})$, the intersection of
$W_+(2)$ and $W_-(2)$ is exactly the set of linearly dependent
$(\xi_1,\xi_2)$'s in $W(2)$. (Cf.\ {\sc Figure} 4-4.)

\bigskip

\begin{flushleft}
{\bf How $W(2)$ embeds in $M(2,{\Bbb C})$.}
\end{flushleft}
Recall from Sec.\ 1 the Milnor fibration $\varphi$ from $S^7-K_{\Xi}$
to $S^1$ for dimension two. The determinant function $\det$ is now a
nondegenerate quadratic polynomial in four variables and hence $\Xi$ has
only an isolated $A_1$-singularity at the origin $O$. The Milnor fiber
$\mbox{\it SL}\,(2,{\Bbb C})$ is homeomorphic to the tangent bundle
$T_{\ast}S^3=S^3\times{\Bbb R}^3$. Its boundary in $S^7$ is $K_{\Xi}$,
which is homeomorphic to the unit tangent bundle $T_1S^3=S^3\times S^2$.
The union
$\varphi^{-1}(e^{i\theta})\cup K_{\Xi}\cup\varphi^{-1}(-e^{i\theta})$
for every $\theta$ is a smooth $S^3\times S^3$ embedded in $S^7$.
More explicitly, if one lets $\Delta$ be the diagonal of $S^3\times S^3$
parametrized by $(p,p)$, $p\in S^{n-1}$, and $\overline{\Delta}$ be the
anti-diagonal parametrized by $(p,\overline{p})$, where $\overline{p}$ is
the antipodal point of $p$, then $\varphi^{-1}(e^{i\theta})$ forms the
$\frac{\pi}{2}$-neighborhood of $\Delta$ while
$\varphi^{-1}(-e^{i\theta})$ forms the $\frac{\pi}{2}$-neighborhood of
$\overline{\Delta}$, where the radius $\frac{\pi}{2}$ is measured at a
$(p,p)$ or $(p,\overline{p})$, either horizontally or vertically in
$S^3\times S^3$, from the latitude in $S^3$ that takes $p$-$\overline{p}$
as north-south poles {\sc Figure 4-1}. All these either are or follow from
well-known ([Di2]) properties of $A_1$-singularities and $S^3$.
Thus, from Sec.\ 1, $K_W$ lies in
$\varphi^{-1}(\{\pm 1, \pm i\})\cup K_{\Xi}$, which is the union of two
$S^3\times S^3$ pasted along $K_{\Xi}$ ({\sc Figure} 4-1, 4-4(b)). The
$S^3\times S^3$ associated $\varphi^{-1}(\{\pm 1\})$ contains the base of
the cone $W_+(2)$ while the $S^3\times S^3$, associated to
$\varphi^{-1}(\{\pm i\})$, contains the base of the cone $W_-(2)$. 

\begin{figure}[htbp]
\setcaption{{\sc Figure 4-1.}
\baselineskip 14pt  
 How $\varphi^{-1}(e^{i\theta})$, $\varphi^{-1}(-e^{i\theta})$, and
 $K_{\Xi}$, for a given $\theta$ occupy the product $S^3\times S^3$ is
 illustrated. The set $\varphi^{-1}(e^{i\theta})$ is indicated by the
 shaded part, a $\frac{\pi}{2}$-neighborhood of the diagonal.   }
\centerline{
   \psfig{figure=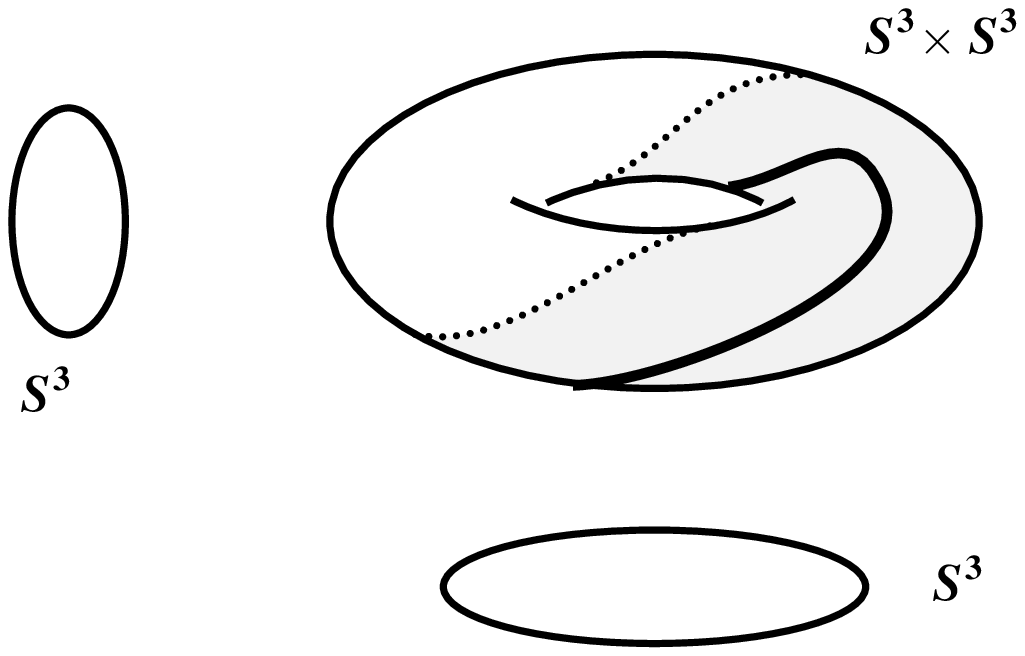,width=13cm,caption=}}
\end{figure}

In this particular dimension, there is another way to see how $W(2)$
embeds in $M(2,{\Bbb C})$, using the fact ([Ro]) that
$$
 S^{p+q+1}\; =\; S^p \ast S^q\,,
$$
where we recall that the {\it join} $X\ast Y$ of two topological spaces
$X$ and $Y$ is defined to be the space $X\times Y\times [0,1]$ with the
following identifications:
\begin{quote}
 \parbox{2em}{(i)}
 \parbox{10cm}{
  $(x,y,0)\sim(x,y^{\prime},0)$ for all $x \in X$,
  $y, y^{\prime} \in Y$; and   }

 \parbox{2em}{(ii)}
 \parbox{10cm}{
  $(x,y,1)\sim(x^{\prime},y,1)$ for all $x, x^{\prime} \in X$,
  $y \in Y$.   }
\end{quote}
The two generating spheres, $S^p$ and $S^q$, of the join is embedded in
the resulting $S^{p+q+1}$ with linking number $1$ and this gives a
fibration of $S^{p+q+1}$ over $[0,1]$ with generic fiber $S^p\times S^q$,
which collapses to $S^p$ over $\{0\}$ and to $S^q$ over $\{1\}$.
({\sc Figure 4-2}). Let us now employ this to our problem.

\begin{figure}[htbp]
\setcaption{{\sc Figure 4-2.}
\baselineskip 14pt  
  The realization of $S^{p+q+1}$ as the join $S^p\ast S^q$
  (cf.\ [Ro], p.\ 5) and the associated fibration of $S^{p+q+1}$ over
  $[0,1]$ (cf.\ {\sc Figure} 4-3). In the picture on the left,
  $S^{p+q+1}$ is regarded as ${\Bbb R}^{p+q+1}\cup\{\infty\}$.  }
\centerline{\psfig{figure=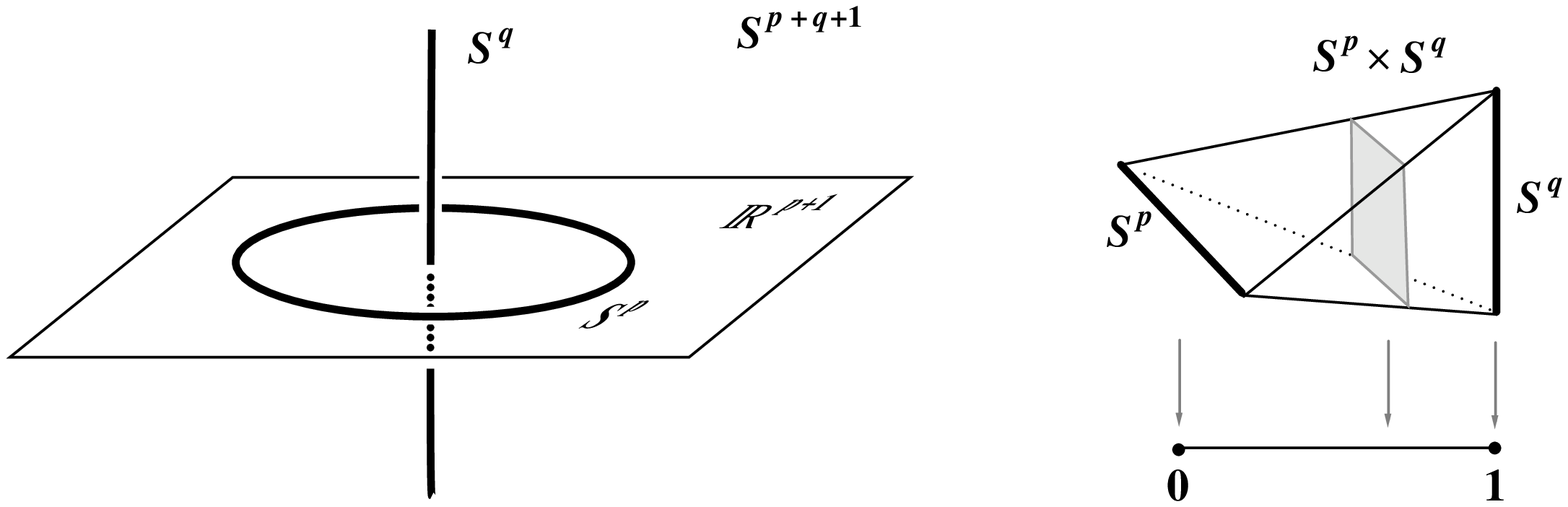,width=13cm,caption=}}
\end{figure}

Up to a slight modification, we assume that $\rho$ is positive-definite.
Hence $V={\Bbb R}^2$ with the standard orthonormal basis $(e_1,e_2)$ and
coordinates $(x_1,x_2)$. We shall adapt ourselves to the following
identifications: $V_{\subscriptsizeBbb C}={\Bbb C}^2={\Bbb R}^4$ with
coordinates
$$
 \xi\; =\;(z_1\,,\,z_2)^t\; =\;(x_1+i\,y_1\,,\,x_2+i\,y_2)^t\,;
$$
and $M(2,{\Bbb C})={\Bbb C}^4={\Bbb R}^8$ with coordinates
$$
 \left(\begin{array}{cc}
  z_1 & w_1 \\ z_2 & w_2
 \end{array}\right)\;
    =\; \left(\begin{array}{cc}
         x_1+i\,y_1 & u_1+i\,v_1 \\ x_2+i\,y_2 & u_2+i\,v_2
        \end{array}\right)\;
    =\; \left(\begin{array}{cccc}
         x_1 & y_1 & u_1 & v_1 \\ x_2 & y_2 & u_2 & v_2
        \end{array}\right)\,.
$$
We also regard $M(2,{\Bbb C})$ as
$V_{\subscriptsizeBbb C}\oplus V_{\subscriptsizeBbb C}$ with metric
$\overline{\rho}\oplus\overline{\rho}$. We shall identify the $S^7$ with
the unit sphere at the origin in any of the above identifications.

In terms of these coordinates, $K_W$ is described by the following system:
$$
\begin{array}{c}
 x_1 y_1\,+\,x_2 y_2\,=\,0\,,\hspace{1em}
                          u_1 v_1\,+\,u_2 v_2\,=\,0\,   \\[1ex]
   x_1 v_1\,+\,y_1 u_1\,+\,x_2 v_2\,+\, y_2 u_2\,=\,0\, \\[1ex]
 x_1^2\,+\,y_1^2\,+\,x_2^2\,+\,y_2^2\,+\,
     u_1^2\,+\,v_1^2\,+\,u_2^2\,+\,v_2^2\,=\,1\,.
\end{array}
$$
Perform now the following change of coordinates, which is an isometry
with respect to $\overline{\rho}\oplus\overline{\rho}$,
$$
 \begin{array}{l}
  x_1\,=\,\frac{\hspace{1ex}\,\overline{x}_1\,-\,\overline{x}_2\,
                +\,\overline{u}_1\,+\,\overline{u}_2}{2}\,,\\[1ex]
  y_1\,=\,\frac{-\overline{x}_1\,-\,\overline{x}_2\,
                -\,\overline{u}_1\,+\,\overline{u}_2}{2}\,,\\[1ex]
  x_2\,=\,\frac{\hspace{1ex}\,\overline{y}_1\,-\,\overline{y}_2\,
                +\,\overline{v}_1\,+\,\overline{v}_2}{2}\,,\\[1ex]
  y_2\,=\,\frac{-\overline{y}_1\,-\,\overline{y}_2\,
                -\,\overline{v}_1\,+\,\overline{v}_2}{2}\,;
 \end{array}
\hspace{3em}
\begin{array}{l}
  u_1\,=\,\frac{\hspace{1ex}\,\overline{x}_1\,+\,\overline{x}_2\,
                -\,\overline{u}_1\,+\,\overline{u}_2}{2}\,,\\[1ex]
  v_1\,=\,\frac{-\overline{x}_1\,+\,\overline{x}_2\,
                +\,\overline{u}_1\,+\,\overline{u}_2}{2}\,,\\[1ex]
  u_2\,=\,\frac{\hspace{1ex}\,\overline{y}_1\,+\,\overline{y}_2\,
                -\,\overline{v}_1\,+\,\overline{v}_2}{2}\,,\\[1ex]
  v_2\,=\,\frac{-\overline{y}_1\,+\,\overline{y}_2\,
                +\,\overline{v}_1\,+\,\overline{v}_2}{2}\,,
 \end{array}
$$
the system then becomes: (with the
\raisebox{1ex}{$\overline{\hspace{1ex}}$} removed for simplicity of
notations in all the rest of discussions)
$$
\begin{array}{c}
 x_1^2\,+\,y_1^2\,+\,x_2^2\,+\,y_2^2\,=\,\frac{1}{2}\,, \hspace{2em}
 u_1^2\,+\,v_1^2\,+\,u_2^2\,+\,v_2^2\,=\,\frac{1}{2}\,, \\[1ex]
 x_1 u_1\,+\,y_1 v_1\,+\,x_2 u_2\,+\, y_2 v_2\,=\,0\,,  \\[1ex]
 x_1^2\,+\,y_1^2\,=\,u_2^2\,+\,v_2^2\,,\hspace{2em}
 x_2^2\,+\,y_2^2\,=\,u_1^2\,+\,v_1^2\,.
\end{array}
$$

Let $\widehat{S}^3_r$ be the $3$-sphere
$x_1^2\,+\,y_1^2\,+\,x_2^2\,+\,y_2^2\,=\,r^2$ in ${\Bbb R}^4\times\{0\}$
and $\overline{S}^3_r$ be the $3$-sphere
$u_1^2\,+\,v_1^2\,+\,u_2^2\,+\,v_2^2\,=\,r^2$ in $\{0\}\times{\Bbb R}^4$;
then $S^7=\widehat{S}^3_1\ast\overline{S}^3_1$ with $K_W$ lying in the
generic leaf
$\widehat{S}^3_{\frac{1}{\sqrt{2}}}\times
                         \overline{S}^3_{\frac{1}{\sqrt{2}}}$.
The expression of the system suggests also a fibration of $K_W$ over, say,
$\widehat{S}^3_{\frac{1}{\sqrt{2}}}$. To illuminate this, let us denote
$\widehat{S}^3_{\frac{1}{\sqrt{2}}}$
(resp.\ $\overline{S}^3_{\frac{1}{\sqrt{2}}}$) simply by $\widehat{S}^3$
(reps.\ $\overline{S}^3$) and introduce the coordinates
$(R; r,\theta_1,\theta_2;
   \overline{r},\overline{\theta}_1,\overline{\theta}_2)$
for $S^7$ with
$$
\begin{array}{l}
 x_1\,=\,r\,\cos\theta_1\,,\\[1ex]
 y_1\,=\,r\,\sin\theta_1\,,\\[1ex]
 x_2\,=\,\sqrt{R^2-r^2}\,\cos\theta_2\,,\\[1ex]
 y_2\,=\,\sqrt{R^2-r^2}\,\sin\theta_2\,,
\end{array}
\hspace{3em}
\begin{array}{l}
 u_1\,=\,\overline{r}\,\cos\overline{\theta}_1\,,\\[1ex]
 v_1\,=\,\overline{r}\,\sin\overline{\theta}_1\,,\\[1ex]
 u_2\,
  =\,\sqrt{1-R^2-\overline{r}^2}\,\cos\overline{\theta}_2\,,\\[1ex]
 v_2\,
  =\,\sqrt{1-R^2-\overline{r}^2}\,\sin\overline{\theta}_2\,,
\end{array}
$$
where $R\in[0,1]$, $r\in [0,R]$, $\overline{r}\in [0,\sqrt{1-R^2}]$, and
$\theta_1,\,\theta_2,\,\overline{\theta}_1,\,\overline{\theta}_2\,
 \in [0,2\pi)$. Then
$$
 \widehat{S}^3\,\times\,\overline{S}^3\;
  =\; \{\,R = \frac{1}{\sqrt{2}}\,\}\,.
$$
In $\widehat{S}^3$, the set $\{r=0\}\cup\{r=\frac{1}{\sqrt{2}}\}$
describes a link while all other
$\{r=\mbox{constant}\in(0,\frac{1}{\sqrt{2}})\}$ describes a torus
$\widehat{T}_r$; and similarly for the torus $\overline{T}_{\overline{r}}$
defined in $\overline{S}^3$. These come simply from the realization of
$S^3$ as the join $S^1\ast S^1$. ({\sc Figure} 4-3.)

\begin{figure}[htbp]
\setcaption{{\sc Figure 4-3.}
\baselineskip 14pt  
 The foliation of $S^3$ by tori with two degenerate leaves that form a
 link (cf.\ [P-R], p.\ 62). In this picture, $S^3$ is realized as
 ${\Bbb R}^3\cup\{\infty\}$ and part of the tori shown is excised for
 clarity.   }
\centerline{\psfig{figure=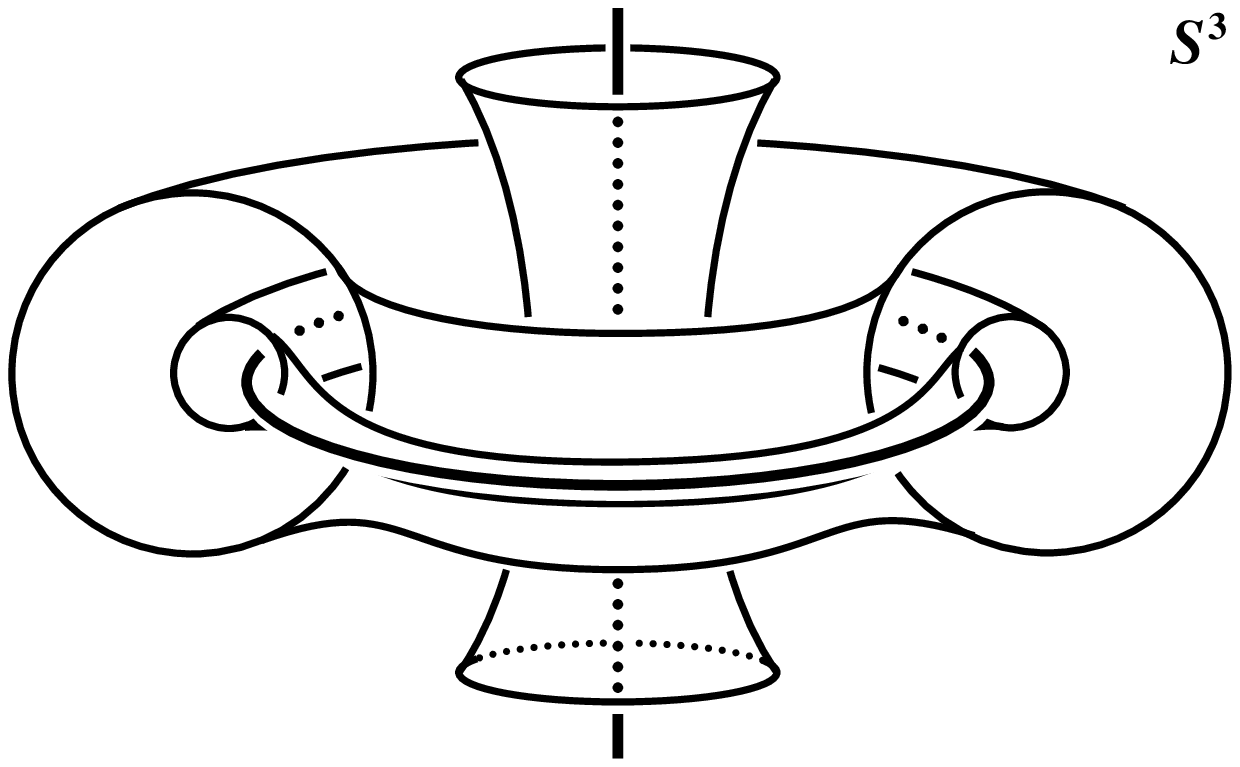,width=13cm,caption=}}
\end{figure}

The system that describes $K_W$ now becomes
$$
 \begin{array}{c}
  R\;=\;\frac{1}{\sqrt{2}}\,, \hspace{1em}
  r^2\,+\,\overline{r}^2\;=\;\frac{1}{2}\,, \\[1ex]
  r\,\overline{r}\,\cos(\frac{(\overline{\theta}_2-\theta_2)
                     + (\overline{\theta}_1-\theta_1)}{2})
 \cos(\frac{(\overline{\theta}_2-\theta_2)
                     - (\overline{\theta}_1-\theta_1)}{2})\;=\;0\,,
 \end{array}
$$
whose solution is given by
$$
\begin{array}{cl}
 \hspace{4em}
  R\,=\,\frac{1}{\sqrt{2}}\,, & \\[1ex]
 \mbox{and} \hspace{3em}
  (r,\overline{r})\,=\,(0,\frac{1}{\sqrt{2}})\,,
      \hspace{2ex}\mbox{\rm or}\hspace{2ex}
                    (\frac{1}{\sqrt{2}},0)\,, & \\[1ex]
 \mbox{\rm or}\hspace{3em}
  (\overline{\theta}_2\,-\,\theta_2)\;
   \equiv\; \hspace{1em}\,(\overline{\theta}_1\,-\,\theta_1)\,+\,\pi
     & (\mbox{\rm mod}\hspace{1ex}2\pi) \\[1ex]
 \mbox{\rm or}\hspace{3em}
  (\overline{\theta}_2\,-\,\theta_2)\;
   \equiv\; -\,(\overline{\theta}_1\,-\,\theta_1)\,+\,\pi
     & (\mbox{\rm mod}\hspace{1ex}2\pi)\,.
\end{array}
$$
Given $(r,\theta_1,\theta_2)$ in $\widehat{S}^3$ with
$r\in(0,\frac{1}{\sqrt{2}})$, the third equation describes the
$(1,1)$-curve $C_-$, and the fouth equation the $(1,-1)$-curve $C_+$,
through $(\theta_1,\theta_2)$ in the torus $\overline{T}_{\overline{r}}$,
where $\overline{r}=\sqrt{\frac{1}{2}-r^2}$ ({\sc Figure 4-4}).
Notice that $C_+$ and $C_-$ intersects also at
$(\theta_1+\pi,\theta_2+\pi)$ in $\overline{T}_{\overline{r}}$. For
$r=0$ or $\frac{1}{\sqrt{2}}$ in $\widehat{S}^3$, the solution is
respectively
the circle $\overline{r}=\frac{1}{\sqrt{2}}$ or $\overline{r}=0$ in
$\overline{S}^3$. Together, this gives a fibration of $K_W$ over
$\widehat{S}^3$ with generic fiber two circles meeting at two points,
which degenerates into a single circle over the link
$\{r=0, r=\frac{1}{\sqrt{2}}\}$ in $\overline{S}^3$.

On the other hand, in terms of
$(R;r,\theta_1,\theta_2;
                \overline{r},\overline{\theta}_1,\overline{\theta}_2)$,
the determinant function $\det$ on $M(2,{\Bbb C})$ restricted to $S^7$ is
given by
\begin{eqnarray*}
\lefteqn{\det|_{S^7} \;
 =\;\left[ r \sqrt{R^2-r^2} \sin(\theta_2-\theta_1)\,
     +\, \overline{r} \sqrt{1-R^2-\overline{r}^2}
       \sin(\overline{\theta}_2-\overline{\theta}_1) \right] }\\[1ex]
 & & \hspace{3em}
  +\, i\, \left[ r \overline{r} \sin(\overline{\theta}_1-\theta_1)\,
     -\, \sqrt{R^2-r^2} \sqrt{1-R^2-\overline{r}^2}
               \sin(\overline{\theta}_2-\theta_2) \right]\,,
\end{eqnarray*}
whose restriction to $K_W$ can be written as 
$$
\begin{array}{c}
 \det|_{K_W}\; =\; r\,\overline{r}\,
             \sin(\frac{\overline{\theta}_2\,-\,\overline{\theta}_1\,
        +\,\theta_2\,-\,\theta_1}{2})\,
      \cos(\frac{(\overline{\theta}_2\,-\,\theta_2)\,
        -\,(\overline{\theta}_1\,-\,\theta_1)}{2})\hspace{6em}\\[1ex]
 \hspace{6em} +\, i\,r\,\overline{r}\,
             \sin(\frac{(\overline{\theta}_1\,-\,\theta_1)\,
                        -\,(\overline{\theta}_2\,-\,\theta_2)}{2})\,
              \cos(\frac{(\overline{\theta}_2\,-\,\theta_2)\,
                        +\,(\overline{\theta}_1\,-\,\theta_1)}{2})\,.
\end{array}
$$

When restricted to $\overline{T}_{\overline{r}}$, observe that the
zero-set of the real part of $\det|_{K_W}$ consists of $C_-$ and the
$(1,1)$-curve
$$
 (\overline{\theta}_2\,-\,\theta_2)\;
  \equiv\; (\overline{\theta}_1\,-\,\theta_1)\,
     -\, 2 (\theta_2-\theta_1)
                     \hspace{1em} (\mbox{\rm mod}\hspace{1ex}2\pi)
$$
while the zero-set of the imaginary part consists of $C_+$ and the
$(1,1)$-curve
$$
 (\overline{\theta}_2\,-\,\theta_2)\;
     \equiv\; (\overline{\theta}_1\,-\,\theta_1)
                      \hspace{1em} (\mbox{\rm mod}\hspace{1ex}2\pi)\,.
$$
This shows that $C_+$ lies in $W_+(2)$ while $C_-$ lies in $W_-(2)$
except for their intersection, whicht lies in $\Xi$. Unless
$\theta_2-\theta_1\equiv\pm\frac{\pi}{2}$ (mod $2\pi$), these loci divide
$\overline{T}_{\overline{r}}$ into six regions, as indicated in
{\sc Figure} 4-4(a). It follows also immediately that the degenerate part
of the fibration of $K_W$ over $\widehat{S}^3$, which are the two
tori in $K_W$ described respectively by $r=0$ and $r=\frac{1}{\sqrt{2}}$,
are in $K_{\Xi}$.

From these data, one observes that, in addition to $C_-\cap C_+$, the
locus $\det|_{K_W}=0$ intersects $C_-\cup C_+$ also at the following
two points
$$
\begin{array}{rcl}
  (\overline{\theta}_1-\theta_1\,,\,\overline{\theta}_2-\theta_2)
   & = & (\,-\,(\theta_1-\theta_2)\,+\,\frac{\pi}{2}\, ,
               \,(\theta_1-\theta_2)\,+\,\frac{\pi}{2}\,) \\[1ex]
   & \mbox{or}
       & (\,-\,(\theta_1-\theta_2)\,+\,\frac{3\pi}{2}\, ,
               \,(\theta_1-\theta_2)\,+\,\frac{3\pi}{2}\,)
                                \mbox{\hspace{1em}(mod $2\pi$)}
\end{array}
$$
in $C_+$. They differ from $C_-\cap C_+$ unless
$\theta_1-\theta_2\equiv \pm\frac{\pi}{2}$ (mod $2\pi$).
In the generic situation when these two points do not coincide with
$C_-\cap C_+$, they correspond to $(\xi_1,\xi_2)$ in $K_W$ that
happens to span a complex line. Following from the indication in
{\sc Figure 4-4}(a), for a generic $(r,\theta_1,\theta_2)$ with
$r\in(0,\frac{1}{\sqrt{2}})$, these four special points decompose
$C_+\cup C_-$ into six arcs. Each lies in a different component of
$K_W-K_{\Xi}$, exhausting the six components of $K_W-K_{\Xi}$. From the
$(\pm,\pm)$ associated to each reagion, where the first $\pm$ indicates
the sign of the real part of $\det$ and the second the sign of the
imaginary part of $\det$, one obtains also how they travel around in the
Milnor fibers, as indicated in {\sc Figure} 4-4(b).

\begin{figure}[htbp]
\setcaption{{\sc Figure 4-4.}
\baselineskip 14pt  
  In (a), the torus $\overline{T}_{\overline{r}}$ is realized as a
  rectangle with the two pairs of parallel edges identified. Its
  coordinates are given in terms $\overline{\theta}_1-\theta_1$
  (horizontal axis) and $\overline{\theta}_2-\theta_2$ (vertical axis)
  that ranges from $0$ to $2\pi$, either rightwards or upwards. The
  locus $\mbox{\rm Re}\,\det|_{K_W}=0$ consists of - - - - - - and $C_-$.
  The locus $\mbox{\rm Im}\,\det|_{K_W}=0$ consists of $\cdots\cdots$
  and $C_-$. In (b), the four Milnor fibers $\varphi^{-1}(1)$,
  $\varphi^{-1}(i)$, $\varphi^{-1}(-1)$, $\varphi^{-1}(-i)$ in $S^7$ that
  contain $K_W$ are indicated and denoted by $F_1$, $F_i$, $F_{-1}$,
  $F_{-i}$ respectively. Their boundary meet at $K_{\Xi}$ (denoted by
  $\Xi_1$ in the picture). The way the generic fiber $C_-\cup C_+$ of
  $K_W$ over $\widehat{S}^3$ travels around these four Milnor fibers is
  shown. The type of the corresponding metrics for each piece is
  indicated, as follows from (a). (Cf.\ {\sc Figure} 3-1.)    }
\centerline{\psfig{figure=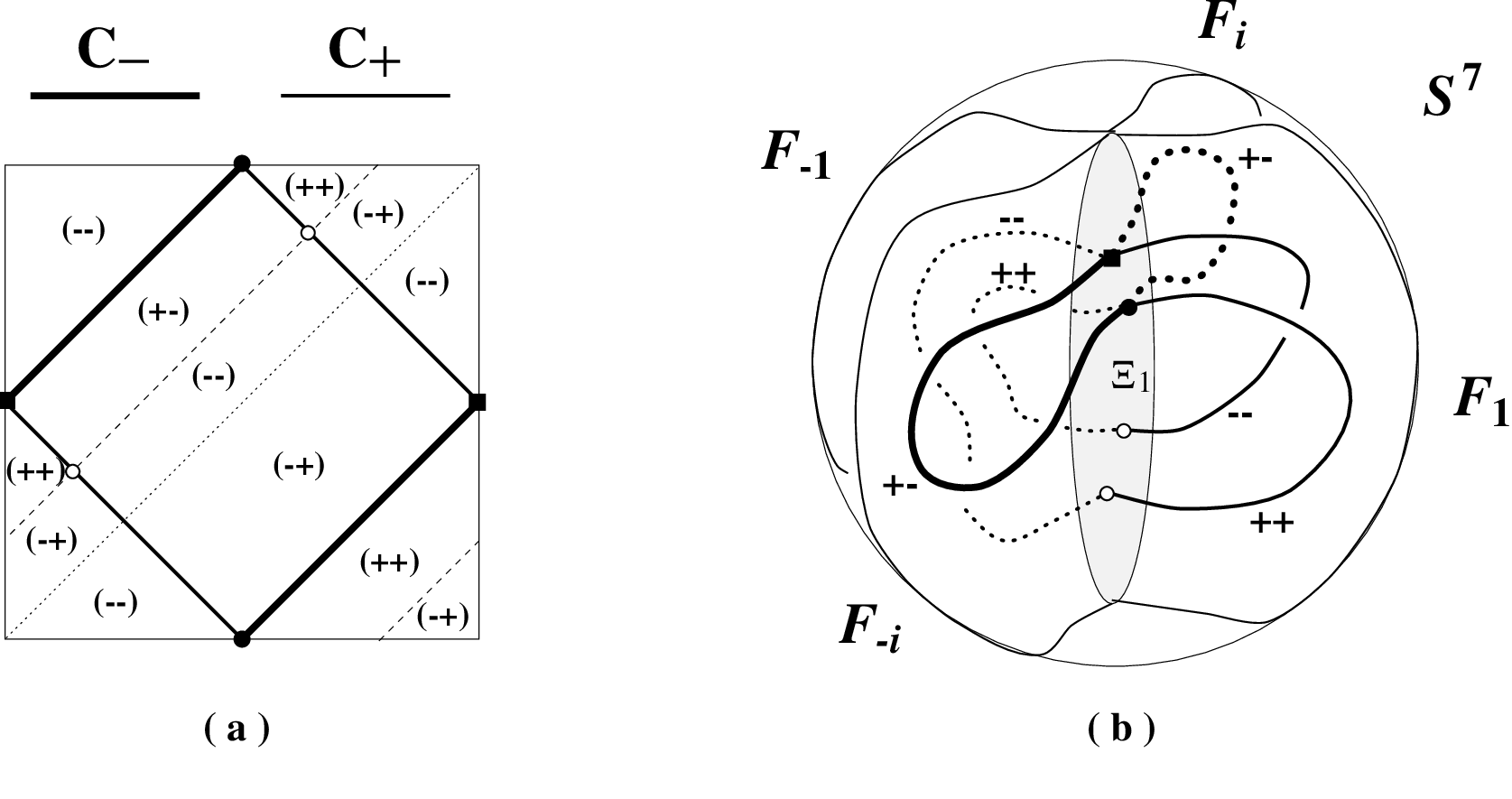,width=13cm,caption=}}
\end{figure}

This concludes our discussion for the embedding of $W(2)$ in
$M(2,{\Bbb C})$.

\bigskip

\begin{flushleft}
{\bf Local Wick rotations of a surface.}
\end{flushleft}
Having said much about Wick rotations, let us now consider local Wick
rotations of a surface $\Sigma$. The following proposition shows that
any generic Lorentzian metric on $\Sigma$ can be inverse-Wick-rotated
to a Riemannian metric. Hence its metric singularities are resolved by
(inverse) local Wick rotations.

\bigskip

\noindent
{\bf Proposition 4.1 [transitivity].}\hspace{.6ex} {\it
 Let $\Sigma=(\Sigma,\rho_0)$ be a compact Riemannian surface. Let
 $\rho_1$ be a generic Lorentzian metric on $\Sigma$. Then there is a
 local Wick rotation
 $f_t:T_{\ast}\Sigma\rightarrow T_{\subscriptsizeBbb C}\Sigma$,
 $t\in[0,1]$, of $\Sigma$ such that
 $f_1^{\ast}{\rho_0}_{\subscriptsizeBbb C}=\rho_1$.
}

\bigskip

\noindent
{\it Proof.} Let $\varsigma=\{p_1, \ldots, p_s\}$ be the set of
singularities of $\rho_1$, where $s=|\chi(\Sigma)|$ is the absolute
value of the Euler characteristic of $\Sigma$, and $\Delta_i$ be a
collection of disjoint disks around $p_i$. From linear algebra, there
exist a pair of line fields $(X_+,X_-)$ on $\Sigma-\varsigma$ that are
simultaneously orthogonal with respect to $\rho_0$ and $\rho_1$ such
that $\rho_1$ is positive-definite on $X_+$ and negative-definite on
$X_-$. Observe that for any $z_+$, $z_-$ non-zero in ${\Bbb C}$,
$z_+ X_+$ and $z_- X_-$ remain orthogonal in
$(T_{\subscriptsizeBbb C}(\Sigma-\varsigma),
                          {\rho_0}_{\subscriptsizeBbb C})$.
Consequently, if one lets
$c_+=\mbox{\raisebox{.6ex}{$\rho_1$}}/
                   \mbox{\raisebox{-.6ex}{$\rho_0$}}$ along $X_+$,
$c_-=-\mbox{\raisebox{.6ex}{$\rho_1$}}/
                   \mbox{\raisebox{-.6ex}{$\rho_0$}}$ along $X_-$,
and $v_+$, $v_-$ be locally-defined nowhere-vanishing vector fields
that lie respectively in $X_+$ and $X_-$, then the family $f_t$ from
$T_{\ast}\Sigma$ to
$T_{\subscriptsizeBbb C}\Sigma$ defined by, for example,
$$
 f_t\;:\; (v_+,v_-)\; \longrightarrow\;
  \left( (1-t+t\sqrt{c_+})\, v_+ \:,
         \:(1-t+t\sqrt{c_-})\,e^{\frac{\pi}{2}it}\, v_- \right)
$$
is singular over $\varsigma$ but is globally well-defined on
$\Sigma-\varsigma$ and is a local Wick rotation from $\rho_0$ to
$\rho_1$ on $\Sigma-\varsigma$. To make $f_t$, $t\ne 1$, an injective
bundle homomorphism from the whole $T_{\ast}\Sigma$ into
$T_{\subscriptsizeBbb C}\Sigma$, one can consider a smooth family of
bump fuctions $\mu^t_i$, $t\in[0,1]$, that take values in $[0,1]$ with
$\mu^t_i(p_i)=1$ and support in $\Delta_i\cap B(p_i,1-t)$, where
$B(p_i,1-t)$ is the ball at $p_i$ of radius $1-t$ with respect to
$\rho_0$. Let $\mu^t=\sum_i \mu^t_i$. One can then redefine $f_t$ by
$\mu^t\,\iota\,+\,(1-\mu^t)\,f_t$, where recall that $\iota$
is the natural inclusion from $T_{\ast}\Sigma$ to
$T_{\subscriptsizeBbb C}\Sigma$. In the limit, we have a bundle
homomorphism $f_1$ from $T_{\ast}\Sigma$ to
$T_{\subscriptsizeBbb C}\Sigma$ that is injective over
$\Sigma-\varsigma$ and is the zero-map at $\varsigma$ such that
$f_1^{\ast}{\rho_0}_{\subscriptsizeBbb C}=\rho_1$. This completes the
proof.

\noindent\hspace{14cm}$\Box$

\bigskip

\noindent
{\it Remark 4.2.}
It is conceivable that, with a careful study of the set of
singularities of generic metrics on a manifold $M$ - it is the set of
singularities of some $p$-plane field on $M$ and has codimension at
least one -, the above assertion and proof should work also at general
dimensions. Also notice that for $M$ an orientable $3$-manifold,
$T_{\ast}M$ is trivial and $M$ admits non-singular metrics of types
$(+,-,-)$ and $(-,+,+)$. Any such metric can be obtained from a local
Wick rotation of a Riemannian one.

\bigskip

\noindent
{\it Remark 4.3.}
Any metrics on $\Sigma$ can be thought of as representatives of their
conformal classes. Hence the above construction gives also a local Wick
rotation from a given Riemann surface to a given Lorentz surface of the
same topology. In [Li] we discussed some details of Lorentz surfaces and
speculated the possible formulation of Lorentzian conformal field theory
(CFT), following Atiyah and Segal's definition. Understanding how the
usual (Riemannian) CFT is transformed under local Wick rotations should
give hints to how Lorentzian CFT can be constructed. This demands works
in the future.

\bigskip

Among the several bundles over $\Sigma$ that appear in the geometry of
local Wick rotations, let us take a look at the $\overline{W}(2)$-bundle,
which concerns the landing of local Wick rotations.

Recall ([Sc], [St]) that an $S^1$-bundle $Y$ over a compact surface
$\Sigma$ is classified by its first Stiefel-Whitney class $w_1(Y)$ in
$H^1(\Sigma;{\Bbb Z}_2)$, if $\partial\Sigma$ is non-empty, and by
$w_1$ and its Euler class $e(Y)$ in $H^2(\Sigma;\{\Bbb Z\})$ if $\Sigma$
is closed, where $\{\Bbb Z\}$ is the local coefficient $\Bbb Z$
on $\Sigma$ that is plainly ${\Bbb Z}$ for $\Sigma$ orientable and is
twisted by the local orientation of $\Sigma$ for $\Sigma$
non-orientable. With respect to a triangulation of $\Sigma$ and up to a
bundle isomorphism that descends to the identity map on $\Sigma$, $w_1$
determines the $S^1$-bundle $Y$ over the 1-skeleton. For $\Sigma$ with
non-empty boundary this is enough to determine the whole bundle. For
$\Sigma$ closed, $e(Y)$ tells in addition how the part of $Y$ over the
2-skeleton is pasted to the part over the 1-skeleton. In terms of
these, we have the following proposition that classifies
$\overline{W}(\Sigma)$.

\bigskip

\noindent{\bf Proposition 4.4.} {\it Given a Riemannian surface
 $\Sigma=(\Sigma,\rho)$. The total space of both components of
 $\overline{W}(\Sigma)$, i.e.\ $\overline{W}_+(\Sigma)$ and
 $\overline{W}_-(\Sigma)$, are orientable. As $S^1$-bundles over
 $\Sigma$, their structures are as listed in the following table:

 \medskip

 $$
  \begin{array}{cccccccccc}
   & & & & & & \underline{\mbox{\makebox[3em]{$w_1$}}}   &
               & \underline{\mbox{\makebox[3em]{$e$}}}   & \\[1ex]
   \partial\Sigma\ne\emptyset\;:  & & & \overline{W}_+(\Sigma)   & &
                          & w_1(\Sigma)   & & \mbox{---}   & \\[1ex]
   & & & \overline{W}_-(\Sigma)  & & & w_1(\Sigma)
                                            & & \mbox{---}   & \\[1em]
    \partial\Sigma=\emptyset\;:   & & & \overline{W}_+(\Sigma)   & &
                                      & w_1(\Sigma)   & & 0   & \\[1ex]
   & & & \overline{W}_-(\Sigma) & & & w_1(\Sigma)   & & 2\, e(\Sigma) &,
  \end{array}
 $$

 \medskip

 \noindent
 where $w_1(\Sigma)=w_1(T_{\ast}\Sigma)$ (vanishes if $\Sigma$ is
 orientable) and $e(\Sigma)=e(T_{\ast}\Sigma)$ are respectively the
 Stiefel-Whitney and the Euler class of $\Sigma$. }

\bigskip

\noindent{\it Proof.}
First it is straightforward to check that the orientability of the
fibre of each bundle along any loop coincides with the orientability of
$\Sigma$ along the same loop. This gives the $w_1$-column. It implies
also that the total space of $\overline{W}(\Sigma)$ are orientable.

For $\Sigma$ closed orientable,
$\overline{W}_+(\Sigma)\cong\Sigma\times S^1$; thus
$e(\overline{W}_+(\Sigma))=0$. As for $\overline{W}_-(\Sigma)$, since
$w_1(\overline{W}_-(\Sigma))=w_1(\Sigma)=0$ from the previous
discussion, one can fix a triangulation of $\Sigma$ and a trivialization
of $\overline{W}_-(\Sigma)$ over the 1-skeleton. Let $T_1\Sigma$ be the
unit tangent bundle of $\Sigma$. As $\mbox{\it SO}\,(2)$-bundles, the
transition function for $T_1\Sigma$ is by left multiplication while
that for $\overline{W}_-(\Sigma)$ is by conjugation. The following
comparison of the two
\begin{eqnarray*}
 \lefteqn{ \hspace{-6em} \left( \begin{array}{rr}
                    \cos\alpha   & -\sin\alpha \\
                       \sin\alpha   & \cos\alpha
                  \end{array}     \right)
           \left( \begin{array}{rr}
                    \cos\theta_0   & \sin\theta_0 \\
                       \sin\theta_0   & -\cos\theta_0
                  \end{array}     \right)
           \left( \begin{array}{rr}
                    \cos\alpha   & \sin\alpha \\
                   -\sin\alpha   & \cos\alpha
                  \end{array}     \right)         }\\
 & & \hspace{-3em} =\;  \left( \begin{array}{rr}
             \cos(2\alpha+\theta_0)   & \sin(2\alpha+\theta_0)  \\
               \sin(2\alpha+\theta_0)   & -\cos(2\alpha+\theta_0)
                 \end{array}   \right)\,,
\end{eqnarray*}
\noindent
while
$$
 \left( \begin{array}{rr}
            \cos\alpha   & -\sin\alpha\\
            \sin\alpha   &  \cos\alpha
        \end{array}    \right)
 \left( \begin{array}{rr}
            \cos\theta_0   &  \sin\theta_0\\
            \sin\theta_0   & -\cos\theta_0
        \end{array}    \right) \;
 =\; \left( \begin{array}{rr}
              \cos(\alpha+\theta_0)   &  \sin(\alpha+\theta_0)\\
              \sin(\alpha+\theta_0)   & -\cos(\alpha+\theta_0)
            \end{array}   \right)\,,
$$
tells us that the pasting homomorphism of $W_-(\Sigma)$ over the
2-skeleton to the part over the 1-skeleton winds twice as many as that
for $T_1\Sigma$. Thus $e(\overline{W}_-(\Sigma))=2e(\Sigma)$.

For $\Sigma$ closed non-orientable, consider its orientation cover
$\widetilde{\Sigma}$ with pullback metric $\widetilde{\rho}$.
Then $\overline{W}_{\pm}(\widetilde{\Sigma})$ is the
pullback bundle of $\overline{W}_{\pm}(\Sigma)$ via the covering map.
By the functorial property of the Euler class and the result for
$\Sigma$ orientable, we have also that
$e(\overline{W}_-(\Sigma))=2e(\Sigma)$. This completes the proof.

\noindent\hspace{14cm}$\Box$

\bigskip

\noindent
{\it Remark 4.5.} For an orientable $3$-manifold $M^3$, since
$T_{\ast}M^3$ is trivial, all the $W(3)$-, $\overline{W}$-, etc.\
bundles over $M^3$, that arise from local Wick rotations, are
trivial.

\bigskip

Having discussed some of the geometric aspects of Wick rotations,
one certainly likes to see its feedback to QFT, gravity, and string
theory, from which the notion arises. We shall leave that for future
works.

\enddocument